\documentclass[twocolumn, superscriptaddress, amsmath, amssymb, aps, floatfix, longbibliography]{revtex4-2}

\usepackage{bm}

\usepackage{dsfont}                           
\usepackage[usenames]{color}                  

\usepackage{graphicx}
\usepackage[caption=false]{subfig}
\graphicspath{ {./images/} } 
\usepackage{tikz}
\usetikzlibrary{quantikz}

\usepackage{dcolumn} 

\usepackage{comment}

\usepackage[en-US]{datetime2}                  

\usepackage{hyperref}                          

\newcommand{\lsim}{\mathrel{\raise.3ex\hbox{$<$\kern-.75em\lower1ex\hbox{$\sim$}}}}
\newcommand{\gsim}{\mathrel{\raise.3ex\hbox{$>$\kern-.75em\lower1ex\hbox{$\sim$}}}}

\def\QECCnk[[#1,#2]]{[\![#1, #2]\!]}

\def\QECCnkq[[#1,#2,#3]]{[\![#1, #2]\!]_{#3}^{\vphantom{T}}}

\def\QECCnkd[[#1,#2,#3]]{[\![#1, #2, #3]\!]}

\def\QECCnkdq[[#1,#2,#3,#4]]{[\![#1, #2, #3]\!]_{#4}^{\vphantom{T}}}

\def\QECCnkgd[[#1,#2,#3,#4]]{[\![#1, #2, #3, #4]\!]}

\def\QECCnkgdq[[#1,#2,#3,#4,#5]]{[\![#1, #2, #3, #4]\!]_{#5}^{\vphantom{T}}}

\def\QECCnkdc[[#1,#2,#3,#4]]{[\![#1, #2, #3; #4]\!]}

\def\QECCnkdcq[[#1,#2,#3,#4,#5]]{[\![#1, #2, #3; #4]\!]_{#5}^{\vphantom{T}}}

\def\QECCnkgdcq[[#1,#2,#3,#4,#5,#6]]{%
  [\![#1, #2, #3, #4; #5]\!]_{#6}^{\vphantom{T}}}

\def\openone{\leavevmode\hbox{\small1\normalsize\kern-.33em1}}

\long\def\symbolfootnote[#1]#2{\begingroup%
\def\thefootnote{\fnsymbol{footnote}}\footnote[#1]{#2}\endgroup}

\newcommand{\be}{\begin{equation}}
\newcommand{\ee}{\end{equation}}
\newcommand{\bp}{\begin{pmatrix}}
\newcommand{\ep}{\end{pmatrix}}
\newcommand{\ben}{\begin{enumerate}}
\newcommand{\een}{\end{enumerate}}

\newcommand{\MS}{M{\o}lmer-S{\o}rensen}

\newcommand{\state}[3]{\textsuperscript{#1}#2\textsubscript{#3}}

\newcommand{\Yb}{\textsuperscript{171}\textrm{Yb}\textsuperscript{+}}

\newcommand{\Htwo}{\ensuremath{\textrm{H}_2}}
\newcommand{\HeHplus}{\ensuremath{\textrm{HeH}^+}}

\begin{document}


\title{Error Mitigation, Optimization, and Extrapolation on a Trapped Ion Testbed}

\author{Oliver G. Maupin}
\affiliation{Department of Physics and Astronomy, 
             Tufts University, 
             Medford, MA, 02155, USA}

\author{Ashlyn D. Burch}
\affiliation{Sandia National Laboratories,
             Albuquerque, NM, 87185, USA}

\author{Brandon Ruzic}
\affiliation{Sandia National Laboratories,
             Albuquerque, NM, 87185, USA}

\author{Christopher G. \surname{Yale}}
\affiliation{Sandia National Laboratories,
             Albuquerque, NM, 87185, USA}

\author{Antonio Russo}
\affiliation{Sandia National Laboratories,
             Albuquerque, NM, 87185, USA}

\author{Daniel S. Lobser}
\affiliation{Sandia National Laboratories,
             Albuquerque, NM, 87185, USA}

\author{Melissa C. Revelle}
\affiliation{Sandia National Laboratories,
             Albuquerque, NM, 87185, USA}

\author{Matthew N. Chow}
\affiliation{Sandia National Laboratories,
             Albuquerque, NM, 87185, USA}
\affiliation{Department of Physics and Astronomy,
             University of New Mexico,
             Albuquerque, NM, 87131, USA}

\author{Susan M. \surname{Clark}}
\affiliation{Sandia National Laboratories,
             Albuquerque, NM, 87185, USA}

\author{Andrew J. \surname{Landahl}}
\affiliation{Sandia National Laboratories,
             Albuquerque, NM, 87185, USA}
 \affiliation{Department of Physics and Astronomy,
             University of New Mexico,
             Albuquerque, NM, 87131, USA}

\author{Peter J. Love}
\affiliation{Department of Physics and Astronomy, 
             Tufts University, 
             Medford, MA, 02155, USA}
\affiliation{Computational Science Initiative,
             Brookhaven National Laboratory,
             Upton, NY, 11973, USA}




\begin{abstract}
Current noisy intermediate-scale quantum (NISQ) trapped-ion devices are subject to errors 
which can significantly impact the accuracy of calculations if left unchecked. A form of error mitigation called zero noise extrapolation (ZNE) can decrease an algorithm's sensitivity to these errors without increasing the number of required qubits. Here, we explore different methods for integrating this error mitigation technique into the Variational Quantum Eigensolver (VQE) algorithm for calculating the ground state of the $\HeHplus$ molecule at $0.8 \textup{\AA}$ in the presence of experimental noise. Using the Quantum Scientific Computing Open User Testbed (QSCOUT) trapped-ion device, we test three methods of scaling noise for extrapolation: time-stretching the two-qubit gates, scaling the sideband detuning parameter, and inserting two-qubit gate identity operations into the ansatz circuit. We find time-stretching and sideband detuning scaling fail to scale the noise on our particular hardware in a way that can be extrapolated to zero noise. Scaling our noise with global gate identity insertions and extrapolating after variational optimization, we achieve error suppression of $96.8 \%$, resulting in an energy estimate within $-0.004 \pm 0.04$ Hartree of the ground state energy. This is an improvement, but still outside the chemical accuracy threshold of $0.0016$ Hartree.  Our results show  that the efficacy of this error mitigation technique depends on choosing the correct implementation for a given device architecture.
\end{abstract}

\maketitle


%
\section{Introduction}
\label{sec:introduction}
Quantum computers offer the promise of speedup over classical computation for several problems, including factoring and quantum simulation \cite{Shor_1997, Lloyd_1996}. One application of quantum simulation that has been extensively studied is in the field of quantum chemistry \cite{Cao_2019, mcardle2020quantum}, particularly the application of finding the ground state energy of a molecule \cite{Aspuru-Guzik:2005a}. A full-configuration interaction (FCI) approach to this requires storing the molecule's full electronic wavefunction, which has a dimension that scales exponentially with the number of electrons in the molecule. However, in a quantum computer the number of qubits required to represent this problem grows polynomially with the number of particles rather than exponentially.

Advances in quantum hardware have brought us into the so-called Noisy Intermediate Scale Quantum (NISQ) era \cite{Preskill_2018}. NISQ machines today are characterized by a number of qubits ranging from 50 to a few hundred, and are notable because machines of this size begin to eclipse what is considered possible to simulate on existing supercomputers \cite{Boixo_2018, Aaronson_2016, Pednault_2017, Google_supremacy}. However, NISQ devices lack error correction, with two-qubit error rates around \( 0.1\% \) \cite{clark2021engineering, Wright_2019, Rigetti_Specs, Google_supremacy, Kim_2023b}. Such errors can dramatically impact the performance of NISQ algorithms if left unchecked, as they can propagate and compound throughout a circuit's run-time. 

Because of this, there has been a great deal of recent work developing, studying, and applying error mitigation methods to NISQ devices, as summarized in \cite{Cai_2022}. These techniques can improve the accuracy and precision of NISQ algorithms without requiring the large infrastructure needed for fault-tolerant quantum computing. In particular, zero-noise extrapolation (ZNE) stands as a method that does not require any qubit overhead, merely requiring precise control of the hardware and increasing the number of expectation values that must be measured. ZNE can reduce the bias of an expectation value at the cost of measuring multiple expectation values, each in turn requiring a sufficient number of samples or circuit executions to build up enough statistics to reach the desired level of precision. Previous studies have shown that this technique can improve the accuracy of expectation values using superconducting qubits \cite{Temme_2017_shortdepth, Kandala_2019_noisy, Kim_2023b} and trapped ion qubits \cite{Shehab_2019, Foss-Feig_2022}.

Here, we focus on the hurdles of practically executing zero-noise extrapolation on a trapped ion device. We study three different implementations of ZNE using known noise scaling methods: 1) Adjusting the duration and detuning of the entangling gates in our circuit while keeping the amplitude constant, 2) Adjusting the detuning and amplitude of the entangling gates while keeping the duration constant, and 3) Inserting entangling gate identity operations.

\begin{table*}[!ht]
    \centering
    \begin{tabular}{|c|c|c|c|}
        \hline
         & Time Stretch Data & Detuning Data & Gate Insertion Data \\
        \hline
        Two-Qubit Gate Estimated State Fidelity  & 98.62 $\pm$ 0.37 \% & 98.86 $\pm$ 1.11 \% & 98.79 $\pm$ 0.66 \% \\
        \hline
        Single-Qubit Estimated Gate Fidelity  & 99.24 $\pm$ 0.56 \% & 99.42 $\pm$ 0.96\% & 99.75 $\pm$ 0.68 \% \\
        \hline
        Qubit Frequency & \multicolumn{2}{l} \centering {12.643 GHz} \cite{clark2021engineering} &  \\
        \hline
        Counterprop Coherence Time & \multicolumn{2}{l} \centering {22.2 $\pm$ 0.9 ms} \cite{clark2021engineering} & \\
        \hline
         Microwave Coherence Time &  \multicolumn{2}{l} \centering {12.6 $\pm$ 0.3 s} \cite{clark2021engineering} & \\
        \hline
    \end{tabular}
    \caption{Specifications of the QSCOUT device}
    \label{tab:device specifications}
\end{table*}

As these noise scaling methods target the dominant sources of noise in a device, they can be used to improve gate performance through repeated calibration, wherein gate parameters are optimized to maximize gate fidelity or some other quality metric. This approach can greatly reduce the error in a device \cite{Gerster_2022} when quantum process tomography is possible, and we do calibrate our gates before experiments. In this work, we focus on using these methods to gauge the effectiveness of application-centric error mitigation, whereby a method's performance is evaluated based on its accuracy in solving a representative problem from a field of practical interest. This approach quantifies errors by their impact on the result of a specific computation, and attempts to improve results without directly addressing sources of error through improvements to gate fidelities.

For current quantum devices, we cannot accurately model every source of noise, particularly when it comes to their effects on algorithms designed for specific applications. We study error mitigation techniques in this context, building on the Variational Quantum Eigensolver (VQE) algorithm \cite{peruzzo2014variational, mcclean2016theory} to solve the electronic structure problem for Helium Hydride.

Our experimental results are obtained using the Quantum Scientific Computing Open User Testbed (QSCOUT), a trapped-ion qubit system based on Ytterbium (\Yb) ions \cite{clark2021engineering}. We investigate how to best implement ZNE on our particular hardware, for our specific optimization procedure, given a fixed sampling budget. These results give us insight into the relative impact of different kinds of device noise, such as error arising from variance in our gate control parameters or heating due to environmental effects. Moreover, these tests indicate how to best mitigate those sources of error in order to improve the accuracy of our algorithm.

In the remainder of Section \ref{sec:introduction}, we introduce the device and algorithm. In Section \ref{sec:optimization} we discuss how we integrate extrapolation into a classical optimization routine. Then, in Section \ref{sec:noise_model}, we discuss our noise model. The following three Sections \ref{sec:time_stretch}, \ref{sec:sideband detuning Scaling}, \ref{sec:gate insertion} outline different noise scaling methods, and the corresponding results of extrapolation in simulation and experiment. We finish with a discussion in Section \ref{sec:discussion}. 

\subsection{Device Details}
\label{subsec:Device Details}
The experiments reported here were performed on QSCOUT's room temperature system, a trapped ion quantum computer that exploits the hyperfine `clock' states of \Yb\ ions (\state{2}{S}{1/2}~$|$F=0, $m_F =0\rangle$ ($\ket{0}$) and $|$F=1, $m_F=0\rangle$ ($\ket{1}$)). The ions are arranged in a linear chain and have all-to-all connectivity through their vibrational modes. Further details of our device can be found in \cite{clark2021engineering}.

Parameterized single and two-qubit gates on this system are generated via Raman transitions using a pulsed 355 nm laser. The pulsed laser is split and delivered to the ions in a counter-propagating configuration, where on one of the arms, a single-channel acousto-optic modulator (AOM) is used to generate a global beam that covers all the ions, and a multichannel L3Harris AOM on another arm is used to generate a set of individual addressing beams that are each focused on individual ions. Using the experiment hardware `Octet' \cite{clark2021engineering}, two frequency tones can be applied to each channel of the AOMs.

For these experiments, we use a native gate set comprised of a single qubit $X$, single qubit $R_{Z}$ rotation, and two-qubit entangling \MS\ gates, as well as the rotation gate $\sqrt{Y}$ for changing measurement basis (FIG.~\ref{fig:ansatz_circuit}). To perform the single-qubit $X$ gate, two Raman tones are applied to single individual addressing beams. The $R_Z(\theta)$ gate is virtual and acts as a phase shift on all subsequent waveforms for that particular qubit. To perform the two-qubit \MS\ ($MS$) gate, two tones are applied to the individual addressing beams targeting the ions in the gate while another tone is applied to the global beam, thereby generating Raman transitions that have been symmetrically detuned from red and blue sidebands. In our system, this is equivalent to an `$XX$' interaction. The inverse gate (MS\textsuperscript{\textdagger}) is performed by shifting the phase of the waveforms applied to the second ion in the gate by $\pi$ radians, converting the gate to an `$-XX$' interaction.

In Table \ref{tab:device specifications} we list more detailed device specifications for each of the experimental datasets discussed in this work. The fidelity is estimated as an upper bound by performing two $MS(\theta = \frac{\pi}{2}$) gates and measuring the resulting probability of the $\ket{11}$ state. This population is then fit to a Gaussian, and the fidelity is obtained by taking the square root of the peak population.

\subsection{Zero Noise Extrapolation}
\label{subsec:ZNE}

Zero noise extrapolation (ZNE) is an error mitigation technique used to improve the accuracy of expectation values when given access to more measurements \cite{Li_2017, Temme_2017_shortdepth}. Rather than attempting to lower error rates, which requires improvements to hardware or calibration, the error in the quantum circuit is instead deliberately and controllably increased. The thought is that if the amount of noise added to the circuit can be carefully scaled by some multiplicative \textit{scale factor} $c$, then multiple noisy expectation values can be measured at set noise intervals. A function can then be fit to this data and traced backwards to obtain an improved, zero noise estimate for the expectation value at $c = 0$. Numerous techniques have been attempted in the past to scale the noise in quantum circuits \cite{Temme_2017_shortdepth, Li_2017, Kandala_2019_noisy, Shehab_2019, Tiron_2020}, some of which will be discussed in Sections \ref{sec:time_stretch}, \ref{sec:sideband detuning Scaling}, and \ref{sec:gate insertion}.

Zero noise extrapolation comes at the cost of increased sampling overhead. For each zero noise estimate, $m$ sampled estimates must be made, each needing samples to meet a desired precision. The variance in the zero noise estimate necessarily depends upon the variance of each of the sampled estimates, but it also depends on the kind of extrapolation used. In the simplest case, a linear function is fit to the data using least-squares regression or something similar. In the case of a purely depolarizing noise model, the data will be best fit by an exponential decay \cite{Li_2017}. As we generally expect that most noise models will have some amount of depolarizing noise, an exponential extrapolation is a good starting point for most noise scaling techniques. However, other kinds of noise may skew the fit.

It is possible that coherent errors might cause the noise to scale quadratically, or cubically, or as some other polynomial function. In that case, if the noise modelling supports it, a higher degree polynomial could be fit to the sampled estimates instead. A special case, known as Richardson Extrapolation \cite{Richardson_1911}, occurs when the highest possible degree polynomial for a given number of data points is fit (polynomial of degree $m-1$ for $m$ data points). Higher degree polynomials double down on the trade-off inherent to extrapolation, further reducing the bias of our estimate but increasing the variance \cite{Tiron_2020}. In the case of Richardson extrapolation, the number of samples required to keep the variance constant scales exponentially with $m$, the number of sampled estimates. As such, if the noise trend is imprecise, it is less useful to fit higher order polynomials. This turns out to the case in this work, as discussed more in Section \ref{sec:gate insertion}. 

Another important consideration is how many sampled estimates (i.e., distinct noise values) are measured for extrapolation. The fewer data points that are used, the more `overfit' the extrapolation will be, with the fitted curve closely following the data. This is good if the noise scaling is very precise, as the resulting zero noise estimate will have high accuracy and precision. However, if there are fluctuations in the values of sampled estimates that are not directly derived from the noise scaling, or if the noise scaling only increases a portion of the noise present in the device, then fitting to fewer data points will result in more bias. On the other hand, increasing the number of sampled estimates not only requires more samples to maintain the desired precision, but also requires that the noise scaling technique allow for that many different noise values to be implemented on the hardware. For the number of samples we have access to on the hardware, we settle upon using a linear or quadratic extrapolation using 4-5 noisy energy estimates fit using weighted least-squares regression. This offers a balance of not over-fitting our extrapolation without ballooning the required number of samples.

\subsection{Error Analysis}
\label{subsec:Error Analysis}
We perform our error analysis and report our metrics of success as follows. The output of $s$ samples of $M$ Hamiltonian terms is a set of $sM$ variables $p^{(k)}_j$ where $1\leq j\leq M$ and $1\leq k\leq s$. We consider the Hamiltonian terms to label the rows of this array of data and the samples to label the columns. The standard analysis of VQE data involves summing the samples along the rows to obtain an estimate of the expectation value of each term in the Hamiltonian. These estimates are then combined with the coefficients of the Hamiltonian to obtain an estimate of the energy expectation value for the Hamiltonian \cite{mcclean2016theory}. The error on this estimate is obtained by standard error propagation from the uncertainties on the terms. 

In this work, we calculate the standard error on the mean of our estimates of the energy expectation value by instead summing the output samples along the columns as implemented in \cite{Ralli_2021}. That is, we take one sample per Hamiltonian term and consider the distribution of the weighted sum of these samples. The mean of this distribution is the same estimate as in the standard analysis.

We report percentage relative errors $\epsilon$ of our measured expectation values $E_{measured}$ compared to the theoretical exact ground state energy $E_{theory}$ as follows:
\begin{equation}
    \epsilon = 100*\frac{|E_{measured}-E_{theory}|}{E_{theory}}
\end{equation}
with accompanying percentage uncertainty $\sigma$ calculated from the standard error of the mean of our measured expectation values $\sigma_{measured}$:
\begin{equation}
    \sigma = 100*\frac{\sigma_{measured}}{E_{theory}}
\end{equation}

When comparing an extrapolated expectation value to an estimate from an unaltered circuit, we report the error suppression as a percentage:
\begin{equation}
    \delta = 100* \left( 1 - \frac{|E_{extrapolated}-E_{theory}|}{|E_{unaltered}-E_{theory}|} \right)
\end{equation}
where this percentage represents how much of the error has been mitigated or reduced by a given error mitigation technique. It is possible that error mitigation will make the error larger, in which case this percentage will be negative.

\subsection{VQE Algorithm and Ansatz}
\label{subsec:VQE}

We use ZNE to mitigate the noise in finding the ground state energy of the small diatomic molecule \HeHplus\ using VQE as in \cite{jaqal_exemplars}. We first encode the electronic Hamiltonian of the molecule to act on qubits using OpenFermion \cite{OpenFermion} in the STO-3G basis. In the Bravyi-Kitaev encoding, the Hamiltonian is a weighted sum of 27 Pauli terms and acts on 4 qubits. Using the qubit-tapering technique in \cite{bravyi2017tapering}, we can reduce the required number of terms to 9 terms acting on 2 qubits because of the $\mathbb{Z}_{2}$ symmetry of the problem arising from conservation of particle number and fermionic parity. The resulting Hamiltonian is:
\begin{gather}
    \begin{aligned}
        H = c_{0} I + c_{1} Z_{0} + c_{2} X_{0} + c_{3} Z_{1} + c_{4} X_{1} \\
        +  c_{5} Z_{1} Z_{0} + c_{6} Z_{1} X_{0} + c_{7} X_{1} Z_{0} + c_{8} X_{1} X_{0}.
    \end{aligned}
\end{gather}

For our ansatz, we implement the Unitary Coupled Cluster for Singles and Doubles ansatz (UCCSD) \cite{hoffmann1988unitary, bartlett1989alternative, taube2006new, harsha2018difference} which can be written as:
\begin{equation}
    U_{UCCSD} = \textrm{exp}(-i \theta X_{1} Y_{0}).
\end{equation}
This can be decomposed into two $MS$ gates and a parameterized $Z$-rotation by a variational parameter $\theta$ as shown in FIG. \ref{fig:ansatz_circuit} \cite{o2016scalable, Hempel_2018}. The Hartree-Fock state for this Hamiltonian is $|\psi \rangle = |11 \rangle$. We initialize our qubits in this state as input into the rest of the ansatz circuit.

\begin{figure}
    \begin{quantikz}
        \lstick{$\ket{0}$} & \gate{X} & \gate[wires=2]{MS} & \qw & \gate[wires=2]{MS^{\dagger}} & \meter{} \\
        \lstick{$\ket{0}$} & \gate{X} & \qw & \gate{R_Z(\theta)} & \qw & \meter{} \\
    \end{quantikz}
    \caption{The UCCSD ansatz circuit for \HeHplus in the STO-3G basis. The circuit structure remains unchanged across different bond lengths, with only the value of the optimization parameter $\theta$ varying.}
    \label{fig:ansatz_circuit}
\end{figure}

For our classical optimizer, we chose to use the Constrained Optimization by Linear Approximation (COBYLA) gradient descent search algorithm as a part of the SciPy optimize package \cite{COBYLA, SciPy}. For a problem of this scale with a single optimization parameter, this algorithm converges quickly and accurately to the ground state, even in the presence of noise. It is important that the classical optimizer be able to efficiently find the ground state in the presence of finite sampling, so that more of the total sampling budget can be allocated to performing the extrapolation procedure that follows. 

This small quantum chemistry problem serves as a good test for exploring how error mitigation techniques such as ZNE can best be integrated into a VQE optimization routine. The simplicity of the ansatz circuit allows us to implement varied methods of noise scaling without the state decohering too quickly. This next section focuses on exploring different ways to combine ZNE and VQE.


\section{Optimization and Extrapolation}
\label{sec:optimization}

An important aspect of this work is the interplay between error mitigation and the classical optimization routine used in VQE. Extrapolation can increase the accuracy of expectation values measured on a noisy device, but it often does so at the expense of the precision of that estimate, given the same sampling budget. As such, we wanted to know whether it is more efficient to incorporate extrapolation into the optimization routine, or if it is better to save extrapolation for the end of the algorithm to improve the final ground state estimate. This section focuses on answering that question for our given problem using simulations. 

\begin{figure*}
    \subfloat{
        \includegraphics[width=\columnwidth]{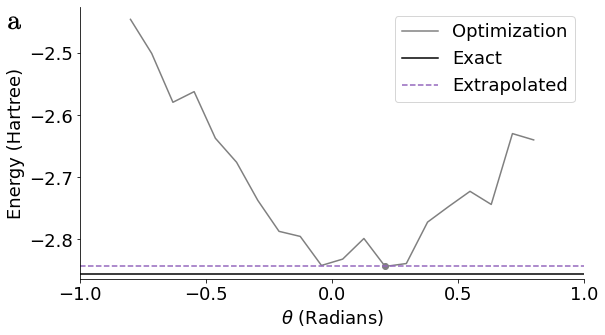}%
        \label{sfig:optimize_over_extrapolated}%
    }
    \hfill
    \subfloat{
        \includegraphics[width=\columnwidth]{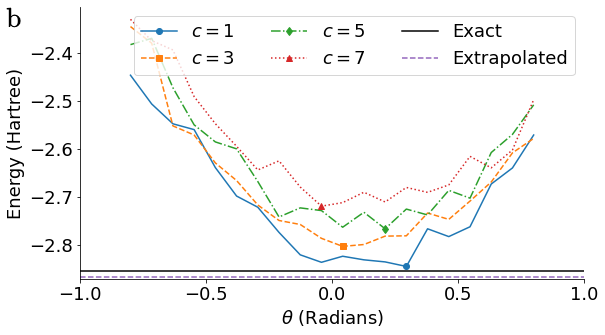}%
        \label{sfig:extrapolate_over_optimized}%
    }
    \\
    \subfloat{
        \includegraphics[width=\columnwidth]{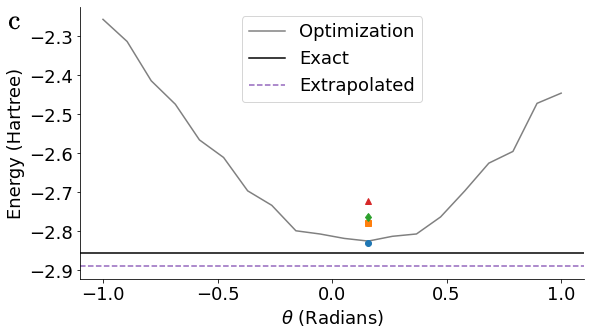}%
        \label{sfig:optimize_then_extrapolate}%
    }
    \hfill
    \subfloat{
        \includegraphics[width=\columnwidth]{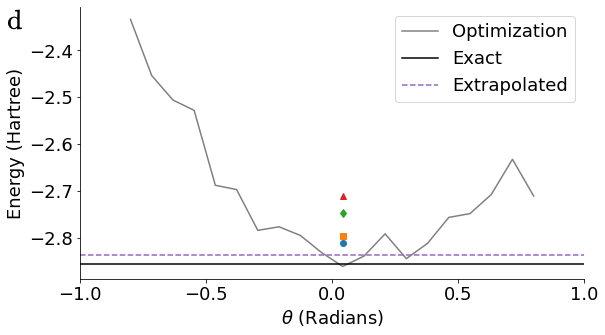}%
        \label{sfig:optimize_then_extrapolate2}%
    }
    \\
    
    \caption{\textbf{Simulation ---} Comparison of four different methods to combine extrapolation and optimization. The single ansatz parameter is divided up into 20 values between $[-1, 1]$, with each routine utilizing the same total budget of $s = 28,000$ samples. The exact ground state energy is shown as a horizontal black line, and the extrapolated zero noise energy estimate as a dashed purple line \textbf{(a)} Extrapolate $m$ energies at each optimization iteration, shown by the solid gray curve. \textbf{(b)} Optimize $m$ separate energy curves, one for each noise scale factor $c$, and extrapolate from $m$ minimum energies (here $m = 4$ highlighted points). In order of increasing noise are the solid blue curve (circle), orange dashed curve (square), dot-dashed green curve (diamond), and dotted red curve (triangle). \textbf{(c)} Find minimum without extrapolation ($m=4$ colored points, same ordering as in \textbf{(b}) and then use remaining sampling budget for final extrapolation. \textbf{(d)} Extrapolate with only two noisy energies at each optimization iteration to find minimum, and then extrapolate at with four noisy energies for final extrapolation.}
    \label{fig:Optimization_Schematic}
\end{figure*}

As was outlined in Section \ref{subsec:ZNE}, the accuracy and precision of extrapolation depends on both the number of noisy energy estimates measured and the number of samples used for those estimates. This sensitivity informs the design of our overall algorithm, as we wish to employ extrapolation and incur its large sampling cost only where it will show improved accuracy over unmitigated estimates. Multiple energy estimates are needed during a VQE optimization, and there are many ways to combine extrapolation and optimization. We now outline four such choices, as shown in FIG. \ref{fig:Optimization_Schematic}.

\textbf{(a) \textit{Optimizing over extrapolated energies}} involves preparing $m$ scaled circuits and measuring $m$ noisy expectation values for a single parameter $\theta$ at each step of the optimization. These estimates are then extrapolated to a single zero noise estimate, which is fed into our optimization routine alongside the parameter $\theta$. This method is more costly, as it requires us to perform our extrapolation numerous times in the middle of our optimization routine. We are no longer optimizing a physical quantum circuit, but rather a mathematical estimation of what we believe the zero noise circuit to be. We note that these extrapolated energies are no longer variational, and therefore can dip below the exact ground state energy. This is a reasonable theoretical approach, however this method potentially exposes our optimization to the sampling sensitivities of higher-order extrapolation. 

\textbf{(b) \textit{Extrapolating over optimized energies}} is the inverse of \textit{optimizing over extrapolated energies}. We prepare $m$ scaled circuits and measure $m$ noisy expectation values of the energy for $m$ optimization parameters $\vec{\theta}$. We then do $m$ independent optimizations in order to determine the minimum noisy energy for each scaled circuit. Once these minima are obtained, we then extrapolate the results to a single zero noise estimate as before. This method is simpler than the first, but poses its own problems. Firstly, it is still computationally intensive, as we must perform $m$ full optimizations. Moreover, depending on how the circuits are scaled, the shape of the energy landscape of the corresponding Hamiltonian may change. This in turn may affect the minimum parameter $\theta$ for each of these scaled circuits. When we extrapolate the $m$ distinct minimum energies, they may correspond to different locations on the estimated energy landscape. We are in effect extrapolating the expectation values of different wavefunctions prepared with different parameters, potentially leading to a less accurate estimate.

\begin{figure}
    \includegraphics[width=\columnwidth]{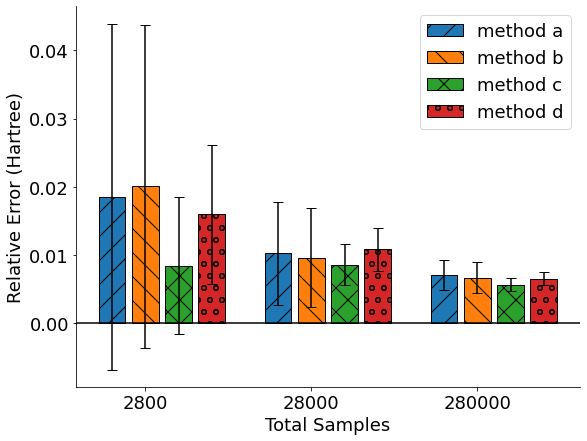}
    \caption{\textbf{Simulation ---} Comparison of the absolute relative accuracy and precision of four different optimization procedures as the total sampling budget is increased. Each method extrapolated the final zero noise energy estimate using a linear fit from $m = 4$ noisy energies using COBYLA optimization to find the optimal parameter $\theta$. The results were averaged across ten trials. \textbf{(a)} Extrapolate from multiple sampled estimates at each step of the optimization routine. \textbf{(b)} Prepare multiple stretched circuits and optimize each of them separately before extrapolating using their respective minima. \textbf{(c)} Optimize an unstretched circuit to find a minimum and then extrapolate. \textbf{(d)} Extrapolate from only two sampled estimates at each step of the optimization, and then perform a high-resolution extrapolation at the minimum. Method c has the lowest variance across the varied number of samples and is the simplest to implement. All estimates fall outside the chemical accuracy threshold of $0.0016$ Hartree.}
    \label{fig:optimization_comparison} 
\end{figure}

\textbf{(c) \textit{First optimizing and then extrapolating}} takes a similar approach to \textit{extrapolating over optimized energies}, though it is further simplified. Instead of optimizing $m$ distinct noisy energy curves, we can instead carry out the full optimization for a single, unmitigated circuit as usual. If we assume that the energy landscape of this base circuit is relatively flat near the minimum, insofar as changes in $\theta$ have a smaller effect on the energy than the noise in our circuit, then this method should give a similar final optimization parameter as the zero noise circuit. We may then prepare $m$ stretched circuits and measure $m$ noisy expectation values of the energy for this final optimized $\theta$ and extrapolate the results to a single zero noise estimate. For a NISQ device like QSCOUT, where we have a finite budget of samples for a given experiment, this third method offers a balance between precision and cost. We find that for a small problem such as \HeHplus with only a single optimization parameter, an unstretched optimization does a good job of finding the minimum, as the energy landscape is a simple sinusoid. Therefore, we can focus our efforts on mitigating the noise present in the final energy estimate rather than spending resources improving the accuracy of our optimization.

\textbf{(d) \textit{Optimizing at low resolution and then extrapolating}} builds on \textit{first optimizing and then extrapolating} similar to what was done in \cite{Kandala_2019_noisy}. If we wish to improve our initial optimization at the cost of more samples, then we can choose to perform our optimization with a low-resolution linear extrapolation procedure at each iteration using the minimum of two data points. This should give us a better noise-free estimate of the optimal variational parameter, and we can then perform a higher-resolution extrapolation using more data points at that parameter to obtain a more accurate final energy estimate. This method serves as a middle ground between methods (a) and (c).

We compare the accuracy and precision of these four algorithms in FIG. \ref{fig:optimization_comparison}, averaged across ten trials. We scaled the noise in our circuits using the \textit{MS Before \& After} method that will be explained in depth in Section \ref{sec:gate insertion}. As expected, the extrapolated energies converge to the ground state energy as we increase the total sampling budget, with slight variations in the accuracy and precision between them. For experiments on the QSCOUT device, we are limited to $\sim 28,000$ total samples for the optimization before qubit-drift starts to become a problem, meaning we can expect relative errors on the order of 1\%.

We find that scaling the noise in the ansatz circuit does not have a large effect on the location of the minimum energy, and therefore we can get a good approximation of the optimal variational parameter by simply optimizing an unmitigated circuit, as evidenced by the accuracy of method c \textit{first optimizing and then extrapolating}. For a more complicated chemical system with a more complex energy landscape, we might have instead chosen method d, \textit{optimizing at low resolution and then extrapolating}, as doing so should mitigate errors incurred while measuring energies during the optimization procedure. This choice ultimately depends on the optimizer's robustness to noise. If VQE has a hard time minimizing the energy of a given Hamiltonian in the presence of errors, then more error mitigation will be needed during optimization. We find that such a technique is unnecessary for \HeHplus.

For the rest of this work, we choose to optimize an unmitigated circuit and then extrapolate at the minimum parameter, corresponding to \textit{first optimizing and then extrapolating} (method c). This choice requires the fewest samples from our total budget for the optimization procedure, meaning that those samples can be instead allocated to obtain precise noisy energy estimates at the optimal variational parameter. This precision will propagate through extrapolation and improve the accuracy and precision of our final zero noise energy estimate of the ground state.


\section{Noise Modeling}
\label{sec:noise_model}
We are interested in both forward and inverse modeling of noise in our device. By forward modeling, we mean a theoretical model of noise that predicts the results of experiment based on our current knowledge of the device. Forward modeling is very important for this work, as it defines how much noise scaling we expect to see for each of the methods we explore, and thus informs our extrapolated estimates. However, we are also interested in inverse modeling through our simulated error model, which is tailored to match current observed properties of the QSCOUT device from calibration, such as gate fidelity.

The following section details the different kinds of errors that our noise modeling takes into account. We only focus on two-qubit gate errors, as our single qubit gates have very high fidelities. We believe that ion-trap devices experience a range of noise, having some sources of time-translation invariant noise, such as that modeled by a depolarizing channel, and some sources of other coherent noise. This is the hypothesis we shall test experimentally in the remainder of this work.

As such, we consider three broad categories of errors. We first consider potential errors at the gate level, based on existing theory and modeled as noise channels. We then discuss the errors in our phenomenological noise model, based on known gate parameter errors in our device, as our qubits do not experience much passive decoherence due to the environment. These errors are physically motivated and abstracted from the circuit model, though some can be directly tied back to noise channels. Lastly are all the errors thought to be present in the device but not captured by our noise model, such as non-Markovian effects or context-dependent errors which are difficult to quantify. These errors remain as roadblocks to fully mitigate error in our experiments.

\subsection{Noise Channels}
\label{subsec:Noise Channels}
An MS gate that rotates by angle $\theta$ about axis $\phi = 0$ acting on two qubits can be written:
\begin{equation}
    MS = \textrm{exp} \left[ -i \frac{\theta}{2} X \otimes X \right]
\end{equation}

During normal device operation, this gate experiences a variety of errors. These errors can be modeled as noise channels acting on the qubit states alongside this `ideal' $XX$ interaction. The gates in our device experience depolarizing noise due to the ions interacting with their environment (i.e., heating or electromagnetic fields in the trap):
\begin{equation}
\label{eq:Depolarizing noise channel}
    \mathcal{E}_{depolarizing} (\rho) = (1-p) \rho + p \frac{I}{2}
\end{equation}
where the qubit state $\rho$ is left unchanged with probability $1-p$ or is transformed into the maximally mixed state with probability $p$. This kind of error is most easily mitigated by ZNE, and is assumed to be present in all experiments in varying strengths.

The gate might also experience crosstalk, wherein the $XX$ operation intended for qubits 0 and 1 bleeds over to act on the extraneous qubit 2:
\begin{equation}
    \mathcal{E}_{crosstalk} (\rho) = \sum_{i=0}^{3} p_{i} E_{i} \rho E_{i}^{\dagger}
\end{equation}
with Kraus operators:
\begin{gather}
    \begin{aligned}
        E_{0} = {(XX)}_{01} \otimes I_{2}, \hspace{5mm} E_{1} = {(XX)}_{02} \otimes I_{1}, \hspace{5mm} \\
        E_{2} = I_{2} \otimes {(XX)}_{01}, \hspace{5mm} E_{3} = I_{1} \otimes {(XX)}_{02}.
    \end{aligned}
\end{gather}
where $\rho$ is the density matrix corresponding to the combined state of all three qubits. Each of these Kraus operators occurs with corresponding probability $p_{i}$. In experiment, we find that the magnitude of our crosstalk error can be time-dependent, which makes it difficult to properly mitigate.

We may also consider other kinds of errors that arise from variations in our laser parameters. One such error in our noise model is the expected variation of the laser power parameter, modeled as a consistent $5\%$ increase from the typical value. Such an increase manifests as an over-rotation by an angle $\delta$ in the MS gate with Kraus operators:
\begin{gather}
    \begin{aligned}
        E_{0} = \textrm{exp} \left[ -i \frac{\theta}{2} X \otimes X \right] = MS \left( \frac{\theta}{2} \right), \\
        E_{1} = \textrm{exp} \left[ -i \left( \frac{\theta}{2} + \delta \right) X \otimes X \right]  = MS \left( \frac{\theta}{2} + \delta \right).
    \end{aligned}
\end{gather}

Each noise scaling method discussed in the remainder of this work will target a different kind of noise. The time-stretching or sideband detuning scaling methods outlined in Sections \ref{sec:time_stretch} and \ref{sec:sideband detuning Scaling} are most effective when applied to a pure depolarizing channel, and may not scale coherent control errors, or may only partially scale them. The gate insertion method in Section \ref{sec:gate insertion} is a bit more general, and can be applied to a wide variety of noise including depolarizing noise and certain kinds of coherent gate errors, but may be ineffective for others such as over-rotation. This discussion is expanded in those later sections.

\subsection{The IonSim Model}
\label{subsec:The IonSim Model}

Throughout this paper, we model single-qubit rotations about the $z$-axis on the Bloch sphere as ideal gates since these are implemented virtually through phase updates on the hardware \cite{clark2021engineering}, and we simulate all other single-qubit gates, $R(\phi, \theta)$, using a square pulse with gate duration of $\tau=(\theta/\pi)\times 22.8\,\mu$s, where $\phi$ and $\theta$ are the rotation axis and the rotation angle on the Bloch sphere, respectively, producing the following unitary transformation,
\begin{equation}
R(\phi, \theta) = \text{exp}\left[-i\frac{\theta}{2}(\cos\phi\sigma_x+\sin\phi\sigma_y)\right],
\end{equation}
where $\sigma_x$ and $\sigma_y$ are the $x$ and $y$ Pauli spin matrices, respectively.

To model MS gates in the presence of realistic device noise, we numerically solve the Master equation,
\begin{gather}
\frac{\mathrm{d}\rho}{\mathrm{d}t} = -i[H,\rho] + \mathcal{L}(\rho), \\
\mathcal{L}(\rho) = \sum_{m=1}^2\left\{J_m\rho J_m^\dagger - \frac{1}{2}(J_m^\dagger J_m\rho + \rho J_m^\dagger J_m)\right\},
\end{gather}
In this model, we use the following Hamiltonian,
\begin{align}
    \label{eq:MS Drive Hamiltonian}
    H = \frac{i\Omega(t)}{2}\sum_j & \eta_j\ket{e}\!\!\bra{g}_j
     \left(a e^{-i\delta_r t}+a^\dagger e^{-i\delta_b t}\right) \nonumber\\
    & \times e^{-i(\phi-\pi/2+\gamma_j)} + \text{h.c.},
\end{align}
which is in a rotating frame with respect to the spin and motional degrees of freedom. Here, $\Omega(t)$ is the time-dependent Rabi rate of the carrier transition between the ground ($\ket{g}$) and excited ($\ket{e}$) spin states with splitting $\omega_{eg}$, $j$ indexes the two ions, $\eta_j$ is the Lamb-Dicke parameter of ion $j$ in a single motional mode, and $\delta_b$ and $\delta_r$ are the detunings of two independent Raman laser tones with frequencies close to the blue and red sideband transitions of this mode, respectively. This mode has an angular frequency of $\nu$, and the sideband detunings are $\delta_b = \omega_b - (\omega_{eg} + \nu)$ and $\delta_r = \omega_r - (\omega_{eg} - \nu)$, where $\omega_b$ and $\omega_r$ are the angular frequencies of the blue-detuned and red-detuned Raman laser tones. The phase $\phi$ corresponds to the rotation axis of the $MS(\phi,\theta)$ gate. The ion-dependent phase $\gamma_i$ is zero for the $MS$ gate, and has the values $\gamma_0=0$ and $\gamma_1=\pi$ for the $MS^\dagger$ gate. 

In addition, we describe the initial motion of the ions before each gate as a thermal state in which $\bar{n}$ is the average number of phonons. As such, the initial density matrix is,
\begin{equation}
    \rho(0) = \rho_\text{spin}\sum_n \frac{1}{1+\bar{n}}\left(\frac{\bar{n}}{1+\bar{n}}\right)^n \ket{n}\!\!\bra{n},
\end{equation}
where $\rho_\text{spin}$ describes the initial spin state, and $\ket{n}$ is a Fock state of the motional mode. During the gate, we describe the heating of this mode through the Linblad jump operators, $J_1=\sqrt{\Gamma}a$ and $J_2=\sqrt{\Gamma}a^\dagger$, where $\Gamma$ is the heating rate.

While we use the same single-qubit gate model throughout this paper, we use two different MS-gate implementations: one for the time-stretched noise scaling that we will describe in Section \ref{sec:time_stretch} and a separate one for the discrete noise scaling that we will describe in Section \ref{sec:gate insertion}. For both MS-gate implementations, we use a Gaussian pulse shape described by $\Omega(t)$. This pulse has a duration of $\tau$, is peaked about $t=\tau/2$, has a standard deviation of $z$, and has a peak Rabi rate of $\Omega_0$. We primarily target the lowest energy motional mode of the ion chain, which is the tilt mode in one orthogonal radial direction. This mode has a motional frequency of $\nu/2\pi=1.75\,$MHz, and we apply two Raman laser tones detuned by $\delta_b=\delta$ and $\delta_r=-\delta$ from the blue and red sidebands of this mode, respectively. In this mode, the two ions have Lamb-Dicke parameters of $\eta_0=0.032$ and $\eta_1=-0.032$.

For the discrete-noise-scaling implementation, we choose $\tau=300\,\mu$s, $z=39.8\,\mu$s, $\delta=-19.6\,$kHz, and $\Omega_0=80.2\,$kHz. While we only include the lowest radial tilt mode in our simulations, the motional mode with the next smallest detuning is the center-of-mass mode in the same radial direction, which has a sideband detuning of $\delta_\text{COM}=175\,$kHz. Since $|\eta\Omega_0/\delta| \approx 0.13$ and
$|\eta\Omega/\delta_\text{COM}|\approx 0.016$,
we see that additional motional modes will make small contributions to the gate dynamics, and we consider their contributions to be negligible compared to the uncertainty in other model parameters.

For the time-stretched $MS$ gates, we implement different gates based on the value of a noise-scaling parameter, $c_\tau$, which linearly stretches the gate duration: $\tau=c_\tau \tau_1$. To target the same gate for different values of $c_\tau$, we fix the peak Rabi rate at $\Omega_0=107\,$kHz and linearly scale the other gate parameters: $z=c_\tau z_1$ and $\delta=c_\tau \delta_1$. For these gates, we choose $\tau_1=200\,\mu$s, $z_1=26.5\,\mu$s, and $\delta_1=-34.5\,$kHz. Similar to the previous implementation, these parameters justify our choice to neglect additional motional modes in this implementation, for all values of $c_\tau$ used in this study.




Our MS gate simulations correspond to an `$XX$' interaction, achieving an $MS$ gate with $\phi=0$ and $\theta=\pi/2$ without errors. To simulate an inverse $MS$ gate, we perform similar simulations but introduce an ion-dependent phase shift of $\pi$ ({\it i.e.} $\gamma_0=0$ and $\gamma_1=\pi$) to produce a `$-XX$' interaction and achieve an $MS$ gate with $\phi=\pi$ and $\theta=-\pi/2$ without errors, which we define as $MS^{\dagger}$.


We then construct the IonSim noise model by introducing errors into the $R$ and $MS$ gate simulations. These errors are informed by characterization experiments that measure typical amounts of drift between calibrations (several hours) and result in typical gate fidelities. Throughout this paper, we use the same error model for each gate in a quantum circuit, without including non-Markovian effects like within-circuit drifts, fluctuations, and context-dependent errors. While non-Markovian effects likely occur in the QSCOUT device, our model simply includes some types of Markovian errors that occur in the device between calibrations.

For both the $R$ and $MS$ gates, we introduce an absolute power offset that increases the Rabi rate by 5\% of its peak value at all times. For the $R$ gates, this is the only error that we include in the model, and it simply increases the gate rotation-angle $\theta$ by 5\%. Including this error, the $R(\phi, \pi/2)$ and $R(\phi, \pi)$ gates have fidelities of $99.8\%$ and $99.4\%$, respectively. 

For the $MS$ gate, in addition to the increase in the Rabi rate by 5\% of its peak value at all times, we also introduce a $500\,$Hz error in the motional frequency $\nu$, give the ions an initial temperature that corresponds to $\bar{n}=0.5\,$q (quanta) at the start of each gate, and introduce a heating rate of $\Gamma=600\,$q/s during each gate.
While the heating rate of the tilt mode is much smaller than $600\,$q/s in the QSCOUT device, this rate represents the effect of heating from all motional modes and the effect of other unknown sources of incoherent noise that reduce the fidelity of the $MS$ gate followed by its inverse.
With these errors, simulations of our discrete-noise-scaling implementation of $MS$ gates produce an entanglement fidelity of 98.5\% for both the $MS$ gate and its inverse, separately, and an entanglement fidelity of 98.1\% for the $MS$ gate followed by its inverse, as measured by the fidelity procedure described in Section \ref{subsec:Device Details}. 

\subsection{Other Experimental Noise}
\label{subsec:Other Experimental Noise}
The IonSim noise model still leaves out other known sources of noise that are present in the physical device. In particular, various non-Markovian processes affect the control of our gates, leading to time-dependent changes in our gate parameters. In general, IonSim does not attempt to model any error that changes between device calibrations or correlated errors between multiple qubits such as crosstalk, and focuses only on the independent, gate-level errors arising from variations in control parameters that was described in the previous section.

The QSCOUT device does require frequent recalibration due to noise affecting the stability of the system. A full VQE optimization may take on the order of an hour to complete, depending on how many samples are taken. Over these timescales, the ions in the device can drift spatially out of the individual addressing beams, causing more gate errors. In particular, this kind of event leads to coherent under rotations of gates within circuits. In addition, fluctuations in the radio frequencies used to create the potential well to trap the ions can impact the device's performance. A large enough fluctuation can induce unwanted sideband transitions and interfere with the execution of gates. This error is not taken into account in our noise model, and as such our experiments on the QSCOUT device may differ from our simulations.

With these sources of noise in mind, we now turn to evaluating the effectiveness of various noise scaling techniques for our hardware. Each noise source will be more amenable to different methods of scaling the noise, and the following sections outline three such approaches, contrasting theory, simulation, and experiment to learn more about how to apply these techniques to our device.

\section{Time-Stretched Noise Scaling}
\label{sec:time_stretch}
We now turn to the first of three methods we tested for increasing the noise in our device. We began by assuming our device experiences the simplest noise channel of pure depolarizing noise. This sort of noise is most amenable to extrapolation and is most easily scaled through simple techniques such as increasing the duration of our gates, which will be the focus of this section.

\subsection{Time Stretch Theory}
\label{subsec:Time Stretch Theory}
In general, the strength of the noise in a quantum circuit is not easily controllable through experimental means, as that would require a highly detailed noise model of the device. By definition, noise arises because of effects in the system that are not understood in detail. However, as was first shown in \cite{Kandala_2019_noisy}, this obstacle can be circumvented if we assume the noise in the circuit is time-translation invariant. For solid-state qubits, such as superconducting devices where qubits rest on top of a substrate and are constantly interacting with their environment, this assumption generally holds. The environment acts as a time-independent source of error over the duration of the gates. For other architectures, there may be no such source of time-translation invariant noise, with the vast majority of noise coming from control errors in the implementation of quantum gates. 

As in \cite{Kandala_2019_noisy}, we first implemented ZNE by scaling the time it takes for our circuit to run. Assuming the noise is time-translation invariant, the noisy expectation value of the circuit will be the same if we run it with noise \( \lambda \) for a scaled time \( c_i \tau \) as it would be for scaled noise \( c_i \lambda \) for a time $\tau$. We can make sampled estimates of the energy by lengthening the duration of gates in our circuit, rather than scaling the noise directly. We implement this in the hardware by keeping the amplitude of the control pulse constant while increasing the duration of the pulse and increasing the detuning of the motional sidebands. We will refer to this method as time-stretching the gates.

\subsection{Time Stretch Simulation}
\label{subsec:Time Stretch Simulation}

We begin our investigation of this noise scaling method with simulations using the IonSim noise model. Though these simulations may not match the performance of the device, due to the presence of errors that are not modeled as described in Section \ref{subsec:The IonSim Model}, they are still valuable as a guide for choosing a parameter space to explore on the QSCOUT hardware. These results inform our choices of how many noise scale factors $c_{i}$ we should use in our extrapolation, what the values of those scale factors should be, how many samples we will need to reach our desired precision, and what the general shape of our energy landscape should look like as a function of the noise scaling.

For our simulations, we include a time-stretch factor $c_{\tau}$ indexed from $i=1,2 ... ,m$ that scales the gate duration by a multiplicative factor between 1.0 and 1.6, where $\tau'= \tau c_i$ and $i$ indexes each sampled estimate. In order to achieve ideal $MS$ gate performance in the noise-free model, the Rabi rate $\Omega$ is held constant and the detuning $\delta$ needs to be scaled by $c_i$ as well, $\delta'=c_i * \delta$. Although the time-stretch factor does not modify the magnitude of the physical errors in the IonSim noise model, this factor increases the sensitivity of the gate performance to errors due to the side effects it has on the Rabi rate and detuning.

\begin{figure}
    \subfloat{
        \includegraphics[width=0.95\columnwidth]{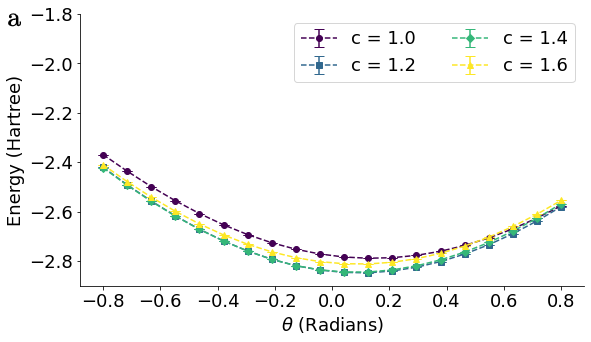}%
        \label{sfig:time_stretch_infinite_simulation}%
    }
    \\
    \subfloat{
      \includegraphics[width=0.95\columnwidth]{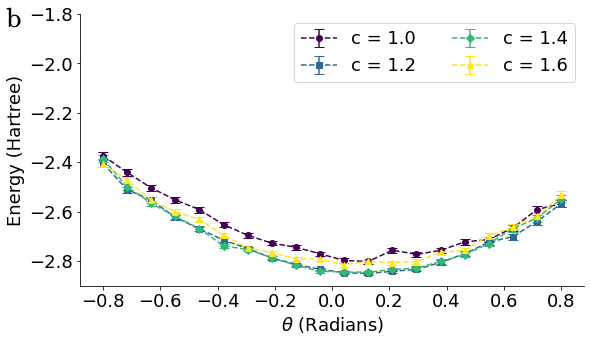}%
      \label{sfig:time_stretch_finite_simulation}%
    }
    \\
    \subfloat{
      \includegraphics[width=0.95\columnwidth]{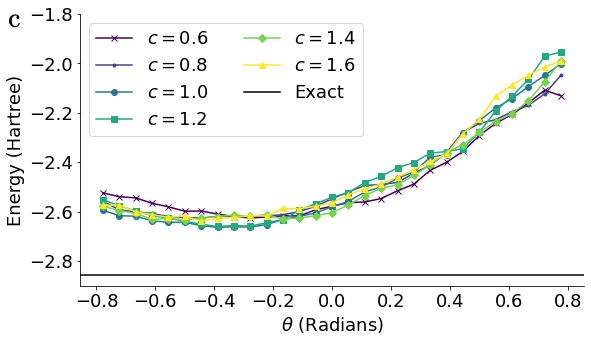}%
      \label{sfig:time_stretch_hardware}%
    }
    \caption{\textbf{Simulation and Experiment ---} Comparison of time-stretched extrapolation. Curves are shown in color, labeled with a stretched duration relative to an unstretched circuit of duration $\tau = 1.0$ . \textbf{(a)} Wavefunction computation of expectation value without finite sampling. The simulated noisy energy curves show some overlap and exhibit non-linear scaling. \textbf{(b)} Simulating 2000 samples for each Pauli term in the Hamiltonian, the energy curves with finite sampling have significant overlap, leading to poor extrapolated estimates. \textbf{(c)} In experiment, gate duration was varied between $0.6$ and $1.6$ times the typical operating duration of $\tau = 200 \mu s$. $2000$ samples were used to estimate each expectation value. There is no increase in energy error as a function of gate duration.}
    \label{fig:time_stretch_simulation}
\end{figure}

In simulation, we find that the error in the final energy estimates does not scale linearly as a function of the noise scale factor $c_{\tau}$, as shown in FIG. \ref{sfig:time_stretch_infinite_simulation}. We find that as we increase the duration and the detuning, the noise first decreases, and then increases as the circuit is lengthened. This is because our noise model was not constrained such that a time stretch factor of $c_{\tau}=1.0$ must correspond to the typical optimal fidelity of the device, and as a result the time stretch/scale factor relationship is offset.

Regardless, the magnitude of the variation in the noise is very small. A variation on the order of $\sim 1-2 \%$ of the unmitigated circuit's minimum energy is introduced by time-stretching our circuits, meaning that the sampled estimates are tightly clustered together. This is in contrast with previous results from \cite{Kandala_2019_noisy}, wherein noise scaling is on the order of $\sim 1-10 \%$. As seen in FIG. \ref{sfig:time_stretch_finite_simulation}, when each energy estimation is made with a fixed sampling budget of 2000 samples, the effect of finite sampling may cause the noisy estimations to further overlap, rendering extrapolation ineffective.

To solve this issue of resolution, we might try to increase the maximum noise scaling of our circuits from 2 up to 5 or even 10 to increase the spacing of the noisy energy curves. However, increasing the duration of the circuit in this way directly scales the runtime as well. In order to get noise scaling of a magnitude amenable to extrapolation, we would need to increase the duration by at least a factor of 10. But, if we scale the noise too much, then the resulting decoherence may cause the noise scaling to become even more nonlinear, or worse render the action of our ansatz circuit useless.

\subsection{Time Stretch Experiment}
\label{subsec:time_stretch_exp}

We began our experiments by similarly testing the effectiveness of time-stretching via a sweep of the energy landscape. We measured our circuits, manually varying the optimization parameter $\theta$ from $-1.0$ to $1.0$ in intervals of $0.05$. The result was a slice of the energy landscape for a given circuit timescale. We then repeated this process, changing the timescale of our $MS$ gates between $0.6$ and $1.6$ times the standard scaling. We refer to this sort of procedure as a `parameter sweep' going forward in this work. The results of this test are shown in FIG. \ref{sfig:time_stretch_hardware}.

We found that while the energy did depend on $\theta$, there was little to no dependence on the scale factor $c_{\tau}$. Lengthening and shortening circuits from normal operating pulse duration conditions did not increase the noise enough to differentiate the noisy energy estimates in the presence of fluctuations due to sampling error and qubit drift. As simulations suggest, we are not able to extrapolate to an accurate zero noise estimate. We also separately tested the effects of stretching the gates at extremely long duration, but found that the noise increased abruptly at around $c_{\tau} = 3$.

We believe that our hardware is robust within this region to this specific implementation of pulse duration scaling for ZNE. It is possible that some other implementation of time-stretching would work, but we chose to shift our attention to other methods of scaling the noise going forward. Though a time-stretching implementation of noise scaling circuits worked well in \cite{Kandala_2017} where a superconducting qubit device was tested, the same appears not to hold true for trapped ion devices such as QSCOUT. In practice, lengthening the timescale of our circuit did not change the amount of noise present in our algorithm at reasonable scale factors.

In the end, our simulations and experimental results both show that a time-stretching implementation does not meaningfully add noise to our system. A simple picture of time translation invariant noise modeled as a depolarizing channel is insufficient to describe our system. As such, we must consider additional sources of noise and alternate methods of increasing the noise in our device in order to successfully perform ZNE.

\section{Motional Mode Detuning Scaling}
\label{sec:sideband detuning Scaling}

In addition to increasing the duration of our gates, we studied the effects of decreasing their duration below the typical, ideal value. As shown in Figure \ref{sfig:time_stretch_hardware}, this had little effect in experiment, but preliminary simulations with shorter durations showed increases in the energy error. For the noise model we used, this is equivalent to decreasing a second gate parameter called the motional mode sideband detuning $(\delta)$. If we take a step back and consider sources of noise modeled by IonSim, we find a number of candidate gate parameters for noise scaling. The duration $\tau$ of the control laser pulses that implement our gates is only one of four possible gate parameters ($\tau, z, \delta, \Omega_{0}$) described in equation \ref{eq:MS Drive Hamiltonian}. In theory, any of these other parameters could be changed in opposition to our gate calibrations in order to increase noise, and this section will focus on changes in the detuning $\delta$.

\subsection{Detuning Theory}
\label{subsec:Detuning Theory}
For the previous time-stretching method, the pulse amplitude was held constant, and the pulse duration was directly varied. In order to ensure that the MS gate performs the correct operation at longer duration, a third parameter called the detuning was re-calibrated and increased. We consider now if this parameter is responsible for the error seen in our unmitigated circuits.

By decreasing the power applied to these sidebands, we recalibrate the system to a smaller detuning and determine how much noise is added in this process by measuring an estimate of the relative infidelity of our two-qubit gates as we change the detuning parameter. For each value of the detuning parameter we want to use, we hold the pulse duration constant, and calibrate the amplitude for ideal gate operation. To approximate the fidelity of that operation, two MS gates are stacked to yield a $\ket{11}$ state; the square root of $\ket{11}$ population then becomes an estimated upper bound for the fidelity of a single two-qubit gate.

\begin{figure}
    \includegraphics[width=\columnwidth]{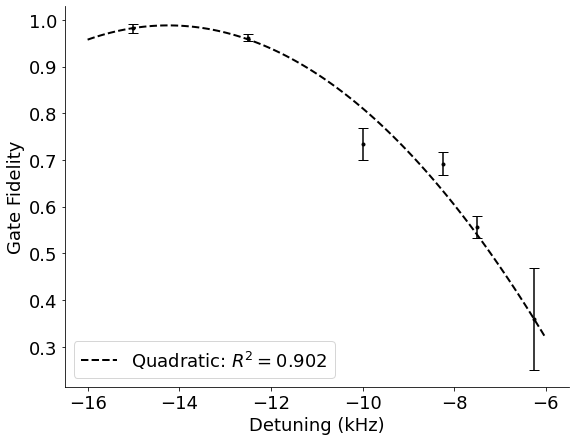}%
    \caption{Experimental measure of two-qubit gate estimated upper bound for fidelity as a function of detuning. The fidelity estimate is highest at a detuning value of $-15 kHz$ and decays quadratically as the detuning parameter increases. Here the fidelity is estimated by running a circuit with two $MS$ gates back-to-back and calculating the square root of the measured probability of the $\ket{11}$ state.}
   \label{fig:detuning_fidelities}
\end{figure}

In practice, we find that there is a range for the detuning parameter that yields the best results, between -15kHz and -7.5kHz. For larger detunings outside this range, the gate fidelity asymptotes to the experimental ideal of $\sim 98.5 \%$, with little change in gate fidelity. On the other side of this range, the gate fidelity decreases as a function of detuning and approaches the fidelity of a fully decohered quantum state. Thus, we set our baseline noise to be $c_{1} = 1.0$ for the -15kHz gate, as this is where gate performance begins to noticeably suffer. A plot of the gate fidelity estimate versus detuning parameter is shown in FIG. \ref{fig:detuning_fidelities}.

Subsequent noise scalings are calculated as:
\begin{equation}
    c_{i} = \frac{1-F_{i}}{1-F_{1}}, \hspace{5mm} i = 1, 2, ..., m
\end{equation}
where $F_{i}$ is the measured fidelity of the gate for the i\textsuperscript{th} detuning parameter, as measured by the fidelity procedure described in Section \ref{subsec:Device Details}. Because of the small infidelity for the ideal detuning $(\sim 1.5 \%)$ the noise scalings for this method are much larger than those used for the other noise scaling methods, running as large as $c_{i} \sim 25$.

We did not perform simulations to test the effectiveness of this noise scaling method for two reasons. Firstly, as described previously, preliminary simulations of the time-stretching method indicated that the sideband detuning may have a larger effect on gate performance. Our simulations of the sideband detuning use the same noise model as our time-stretch simulations, and therefore would not provide us new information to guide our experiments. Secondly, our technique for calculating the noise scale factor is dependent on experimental observations of fidelity. As such, we moved to testing this method on the QSCOUT device.

\subsection{Sideband Detuning Experiment}
\label{subsec:sideband detuning experiment}

\begin{figure}
    \subfloat{%
      \includegraphics[width=\columnwidth]{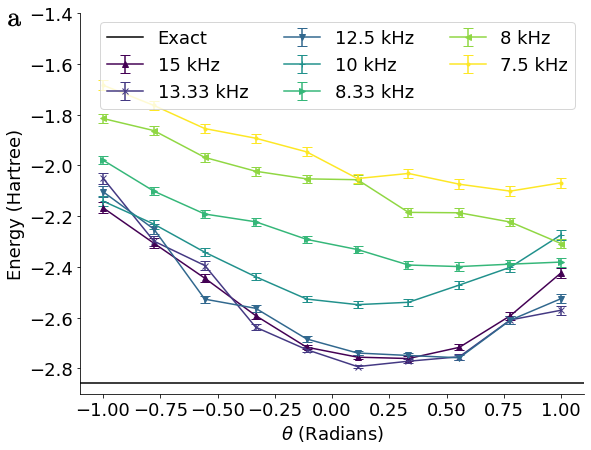}%
      \label{sfig:detuning_parameter_sweep}%
    }\hfill
    \subfloat{%
      \includegraphics[width=\columnwidth]{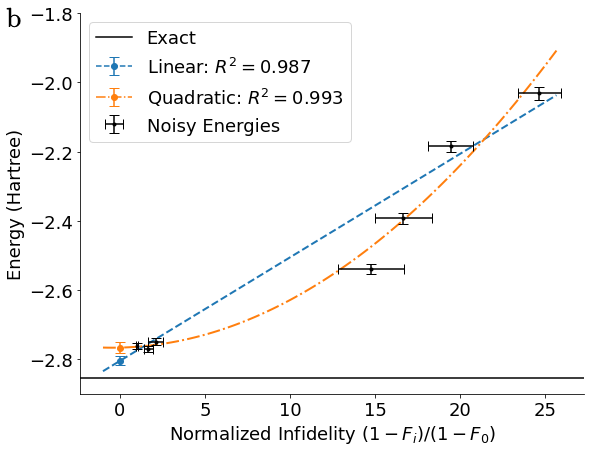}%
      \label{sfig:detuning_extrapolation}%
    }\hfill
    \caption{\textbf{Hardware ---} Experimental implementation of noise scaling by direct scaling of the sideband detuning parameter. \textbf{(a)} Noisy energy estimates were obtained for detunings: $[-15, -13.33, -12.5, -10, -8.33, -8, -7.5]$ kHz, using 2000 samples per projective measurement for different ansatz parameters $\theta$. We see a degradation in the quality of the gate begin around -12.5kHz and level off around -7.5kHz, with a mostly linear regime in-between. \textbf{(b)} The data at the minimum variational parameter $\theta$ was fit using total least-squares to a linear function (blue, dashed) and quadratic function (orange, dot-dashed). The quadratic function best describes the data, leading to a zero noise energy estimate on par with the unaltered circuit. This energy gap implies that this method only scales a portion of the total noise present in the circuit.}
    \label{fig:detuning_parameter_sweep}
\end{figure}

To perform extrapolation using this sideband detuning scaling method, we  adjusted the power applied to the red and blue sidebands such that the necessary detunings for the gate were across a wide range, from -40kHz all the way down to -6kHz. To find each desired detuning, we calibrated the MS gate to specific powers applied to the sidebands and measured noisy expectation values using 2000 samples per projective measurement. Within this range we found that changing the parameter only had an effect on the energy within the range of -15kHz to -7.5kHz, as described in the previous theory discussion and shown in FIG. \ref{sfig:detuning_parameter_sweep}. These energies are plotted versus their corresponding noise scaling in FIG. \ref{sfig:detuning_extrapolation} and fit using total least squares regression from \textsc{SciPy.ODR} \cite{SciPy}. This regression technique takes into account the $y$-error in the noisy energies themselves, as well as the $x$-error derived from our uncertainty of the gate fidelity for different values of the detuning parameter. 

We compare the goodness of our fits by looking at two statistical metrics: reduced chi-squared and adjusted R-squared, computed \textit{ex post facto}. Reduced chi-squared is a squared ratio of how far away each data point is from the fit compared to the error bars of that point, also known as the residual sum-of-squares:
\begin{equation}
    \label{eq:Chi Squared}
    \chi^{2} = SS_{res} = \left( \sum_{i=0}^{n} \frac{(y_{i} - f(y_{i}))^{2}}{\sigma_{y_{i}}^{2}} + \sum_{i=0}^{n} \frac{(x_{i} - f(x_{i}))^{2}}{\sigma_{x_{i}}^{2}} \right)
\end{equation}
Where $n$ is the total number of data points, $x_{i}$ and $y_{i}$ are the $x$ and $y$-positions of the data points respectively, each with error $\sigma_{x_{i}}$ or $\sigma_{y_{i}}$, and $f(x_{i})$ or $f(y_{i})$ are the $x$ and $y$-positions of the fitted curve at the position of the corresponding data point on the opposite axis. (i.e., $(y_{i} - f(y_{i})$ is the usual definition of a residual difference at a fixed point along the $x$-axis).

The reduced chi-squared has an extra factor out front that divides by the number of degrees of freedom in the fit, or the number of data points $n$ minus the number of fit parameters $p$, in order to penalize over-fitting the data:
\begin{equation}
    \label{eq:Reduced Chi Squared}
    \chi_{red}^{2} = \frac{\chi^{2}}{n-p}
\end{equation}

The reduced chi-squared values of the two extrapolations in FIG. \ref{sfig:detuning_extrapolation} are both greater than one, on account of the small y-error bars relative to the difference between the data points and fits. We obtain values of 5.85 and 2.55 respectively for the linear and quadratic fits, indicating the quadratic fit is better.

The second fit metric we use is the adjusted R-squared, which is a measure of how much better a fit is compared to a flat weighted average of the data points. It is calculated by dividing the residual sum-of-squares by the total sum-of-squares, which is a similar measure comparing the difference between each data point and the weighted mean. R-squared is one minus this ratio:
\begin{equation}
        \label{eq:R Squared}
        R^{2} = 1 - \frac{SS_{res}}{SS_{tot}} = \sum_{i=0}^{n} \frac{(y_{i} -f(y_{i}))^{2} + (x_{i} - f(x_{i}))^{2}} {(y_{i} - \boldsymbol{\bar{y}})^{2}}
\end{equation}
where $\boldsymbol{\bar{y}}$ is simply the mean of the data points $\{ y_{i} \}$.

Adjusted R-squared additionally takes into account the number of degrees of freedom in the fit, in a manner similar to the reduced chi-squared metric:
\begin{equation}
    \label{eq:Adjusted R Squared}
    R_{adj}^{2} = 1 - \left( 1 - R^{2} \frac{n - 1}{n - p - 1} \right)
\end{equation}
Comparing the adjusted R-squared values of linear (0.987) and quadratic (0.993) fits to the data, we find that both fits are good on account of the large $x$ error-bars, with a small difference between the resulting zero noise energy estimates. 

Choosing the quadratic fit for extrapolation (orange, dot-dashed) results in a zero noise energy estimate that is very close to that of the first noisy energy estimate, corresponding to the unaltered circuit. This result is slightly worse than the zero noise estimate obtained from the linear fit (blue, dashed), but follows the data more closely. The lack of error suppression implies that this noise scaling method can increase the noise present in our circuit by deliberately mis-calibrating our gates, but that we may not be able to mitigate error with this method, as the resulting noiseless estimate is on par with the performance of an unmitigated circuit. This quadratic trend is similar to that seen in \cite{Rabinovich_2024}, wherein the authors study the effects of stochastic perturbations in gate parameters on the energy error in variational algorithms. Our deliberate changes in the sideband detuning are similar to the magnitude of the random fluctuations in gate parameters in their analysis, leading to similar scaling in energy error.

\definecolor{dark green}{rgb}{0, 0.7, 0}
\definecolor{bright purple}{rgb}{0.75, 0, 0.75}
\begin{figure*}
    \begin{quantikz}
    \lstick{$\ket{0}$} & \gate{X} & \gate[wires=2]{MS^4}\gategroup[wires=2,steps=1,style={dashed, rounded corners, fill=red!60, inner xsep=4pt, inner ysep=2pt}, background]{\textcolor{red}{(d) four}} & \gate[wires=2]{MS} & \gate[wires=2]{MS^{\dagger}}\gategroup[wires=2,steps=5,style={dashed, rounded corners, fill=dark green!60, inner xsep=4pt, inner ysep=2pt}, background]{\textcolor{dark green}{(c) before \& after}}\gategroup[wires=2,steps=2,style={dashed, rounded corners, fill=blue!60, inner xsep=4pt, inner ysep=2pt}, background]{\textcolor{blue}{(a) before}} & \gate[wires=2]{MS} & \qw & \gate[wires=2]{MS^{\dagger}}\gategroup[wires=2,steps=2,style={dashed, rounded corners, fill=orange!60, inner xsep=4pt, inner ysep=2pt}, background]{\textcolor{orange}{(b) after}} & \gate[wires=2]{MS} & \gate[wires=2]{MS^{\dagger}} & \gate[wires=2]{MS^{\dagger4}}\gategroup[wires=2,steps=1,style={dashed, rounded corners, fill=red!60, inner xsep=4pt}, background]{\textcolor{red}{(d) four}} & \meter{} \\
    \lstick{$\ket{0}$} & \gate{X} & \qw & \qw & \qw & \qw & \gate{R_Z(\theta)} & \qw & \qw & \qw & \qw & \meter{}
    \end{quantikz}
    \caption{The UCCSD ansatz circuit for \HeHplus shown in FIG. \ref{fig:ansatz_circuit} with possible gate identity insertions. $(MS^{\dagger} MS)^{i}$ can be inserted before the parameterized $R_{z}(\theta)$ gate \textbf{(a) blue}, after \textbf{(b) orange}, or both \textbf{(c) green}. $(MS)^{4i}$ and $(MS^{\dagger})^{4i}$ can also be inserted in both halves of the circuit \textbf{(d) red}.}
    \label{fig:folded_ansatz_circuit}
\end{figure*}

\section{Gate Insertion Noise Scaling}
\label{sec:gate insertion}
The result of our first two experiments leads us to believe that variation in individual gate parameters is not enough to explain the errors that we see in our unmitigated circuits. A more holistic approach to noise scaling may perform better when we don't have a full picture of noise present in our device. In this section, we outline one such noise scaling method that is more agnostic to noise source.

\subsection{Gate Insertion Theory}
\label{subsec:Gate Insertion Theory}
An alternative manner of scaling control errors in our circuits is to make discrete changes at the circuit-level, rather than changing the duration or detuning of the gates themselves, as described in \cite{Dumitrescu_2018, He_2020, Tiron_2020}. Gate-insertion methods scale the noise in the circuit by varying the depth of the circuit rather than gate duration. A gate may be replaced by a series of gates:
\begin{equation}
    U \rightarrow U (U^{\dagger} U)^{i}
\end{equation}
where $i=1, 2, ... m$ is a positive integer that indexes how many insertions have been made. By inserting the identity $U^{\dagger} U$ this way, the logical structure of the circuit is maintained while the effects of the noise are scaled by a factor $c_{U}$.

This technique can be applied to the entire circuit all at once in a process called \textit{global folding} \cite{Tiron_2020} to get an integer scale factor depending on the number of identity insertions. As we are interested in amplifying the noise due to our $MS$ gates, we limit our insertions to those gates alone using \textit{local folding}:
\begin{equation}
    MS\rightarrow MS(MS^{\dagger} MS)^{i}
\end{equation}
This has the effect of amplifying control errors in our gates in a controlled manner. We tested this identity insertion in three ways: 1. \textit{MS Before} where the identity insertion is placed before the first gate MS in our circuit. 2. \textit{MS After} where the identity insertion is placed after the second MS gate in our circuit. 3. \textit{MS Before \& After} where we do both. These methods are shown in FIG. \ref{fig:folded_ansatz_circuit}.

We note, however, that this method may fail to scale certain kinds of coherent errors in the circuit. Because we are inserting a gate followed by its inverse, it is possible that some errors introduced (such as the over-rotation descibed in Section \ref{subsec:Noise Channels}) will be immediately cancelled out rather than increased. If the majority of the noise in the circuit is due to coherent errors that cancel in this way \cite{Majumder_2023}, then this method will add little to no noise to the circuit. This would render extrapolation futile, as our goal is to amplify all the errors incurred in the unstretched circuit.

A possible workaround to this cancellation is to instead scale the noise by inserting a different set of gates, so long as those gates also comprise an identity operation. For instance, we may choose to instead scale our $MS$ gates as follows:
\begin{equation}
    MS \rightarrow MS (MS)^{4i} \hspace{10mm}  MS^{\dagger} \rightarrow MS^{\dagger} (MS^{\dagger})^{4i}
\end{equation}
where we now insert the identity in the form of four \MS\ gates each with $\phi = 0$, $\theta=\frac{\pi}{2}$, the same parameters as our usual $MS$ gate. This kind of discrete gate insertion would circumvent cancellation of certain coherent errors, as we never follow an entangling operation with its inverse.

However, this method requires that four two-qubit gates be inserted at a time for each preexisting two-qubit gate in our circuit, which introduces a lot of noise very quickly. This low resolution in the amount of noise that we are able to add to the circuit as a function of the number of insertions may make it difficult to perform extrapolation. We want to have enough distinct noisy energy estimates (and thus enough insertions) to ensure that our extrapolation does not over-fit our data, but as we add more and more gates to our circuit we run the risk of decoherence overshadowing all other forms of noise, negating the effect of our ansatz circuit. This method is also shown in FIG. \ref{fig:folded_ansatz_circuit}

We expect all these gate insertion methods (FIG. \ref{fig:folded_ansatz_circuit}) to generally scale the noise in our circuit as a function of the total number of two-qubit gates. It is difficult to model how coherent errors will affect the noise, but if we assume only a depolarizing channel acts on the gates, then the total noise will be increased exponentially with the number of gate insertions, as each additional gate will cause the quantum state to decay:
\begin{equation}
    \mathcal{E} (\rho) = (1 - p^{\lambda}) U \rho U^{\dagger} + p^{\lambda} \frac{I}{2}
\end{equation}
where $\lambda$ is the relative circuit depth of the error-mitigated and unmitigated circuits. That is, the ratio of the number of two-qubit gates in the noisy circuit divided by the number of two-qubit gates in the unmitigated circuit (two here). 

This means that the resulting noise scale factor is also modeled as an exponential in the number of gates added:
\begin{equation}
    c_{U} = a + b e^{\lambda}
\end{equation}
This noise scale factor is normalized such that $c_{U}(\lambda = 0) = 0$ for a zero noise circuit and $c_{U}(\lambda = 1) = 1$ for an unmitigated circuit, which in turn constrains the values of $a$ and $b$. Therefore, we expect our noise scaling to follow the trend:
\begin{equation}
    \label{eq:gate insertion scale factor}
    c_{U} = \frac{e^{\lambda} - 1}{e - 1}
\end{equation}
Deviation from this scaling is a sign that other forms of noise are present in our experiment.

\subsection{Gate Insertion Parameter Sweep}
\label{subsec:Gate Insertion Parameter Sweep}

\begin{figure}
    \subfloat{%
      \includegraphics[width=\columnwidth]{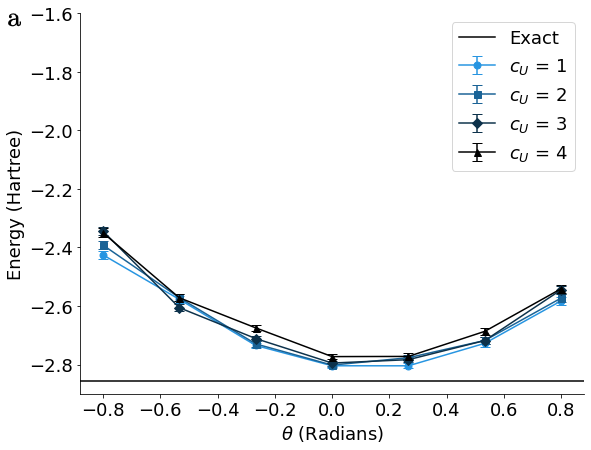}%
      \label{sfig:MS Before Simulation}%
    }\hfill
    \subfloat{%
      \includegraphics[width=\columnwidth]{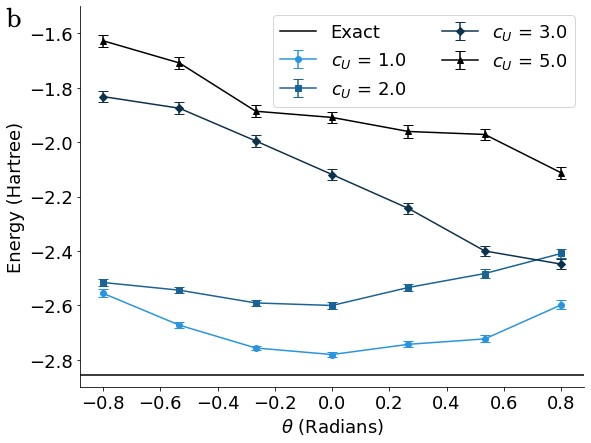}%
      \label{sfig:MS Before Experiment}%
    }\hfill
    \caption{Comparison of simulation \textbf{(a)} and experimental \textbf{(b)} parameter sweep of the \textit{MS Before} method. Gates are inserted into only the first half of the circuit, before the first MS gate. We find that the noise scales linearly as a function of the number of gates in both simulation and experiment. The experimental results show an outlier curve at $c_{U} = 3.0$, likely due to coherent errors.}
    \label{fig:MS Before}
\end{figure}

\begin{figure}
    \subfloat{%
      \includegraphics[width=\columnwidth]{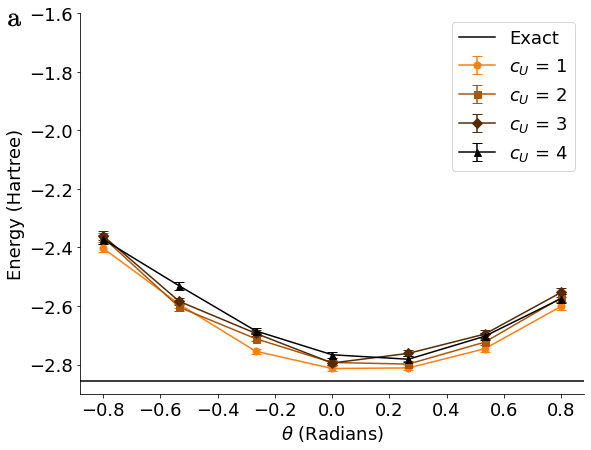}%
      \label{sfig:MS After Simulation}%
    }\hfill
    \subfloat{%
      \includegraphics[width=\columnwidth]{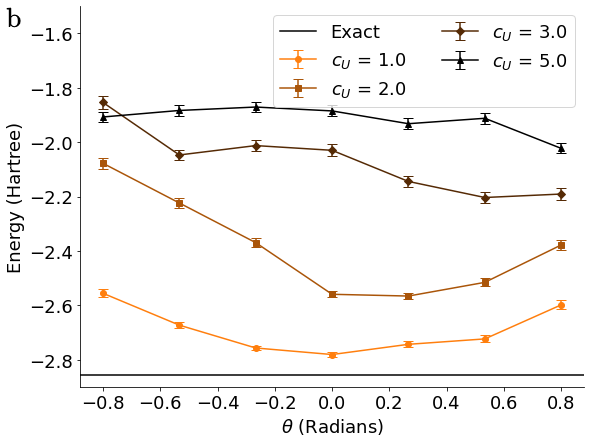}%
      \label{sfig:MS After Experiment}%
    }\hfill
    \caption{Comparison of simulation \textbf{(a)} and experimental \textbf{(b)} parameter sweep of the \textit{MS After} method. Gates are inserted into only the second half of the circuit, after the second MS gate. We find that the noise again scales linearly as a function of the number of gates in both simulation and experiment. The experimental results are more consistent than other gate insertion methods.}
    \label{fig:MS After}
\end{figure}

In this section, we compare the results of both simulated and experimental parameter sweeps for each of the gate insertion techniques shown in FIG. \ref{fig:folded_ansatz_circuit}. We plot noisy energy curves for different scale factors as in Sections \ref{sec:time_stretch} and \ref{sec:sideband detuning Scaling}, varying $\theta$ between $-0.8$ and $0.8$, with $2000$ samples per estimate of the expectation value of each term in the Hamiltonian. For our experiments, the sideband detuning was recalibrated to correct for qubit drift between each trial.

The results broadly show the expected behavior for these four methods, and the noise scaling is improved compared to FIG \ref{sfig:time_stretch_finite_simulation}. That said, the noise scaling does not follow the theoretical exponential decay outlined in the previous section. Instead, the expectation value of the energy follows a roughly linear trend as a function of the number of gates added, signifying that a purely depolarizing model of the noise as described by Equation \ref{eq:gate insertion scale factor} may be insufficient.

\begin{figure}
    \subfloat{%
      \includegraphics[width=\columnwidth]{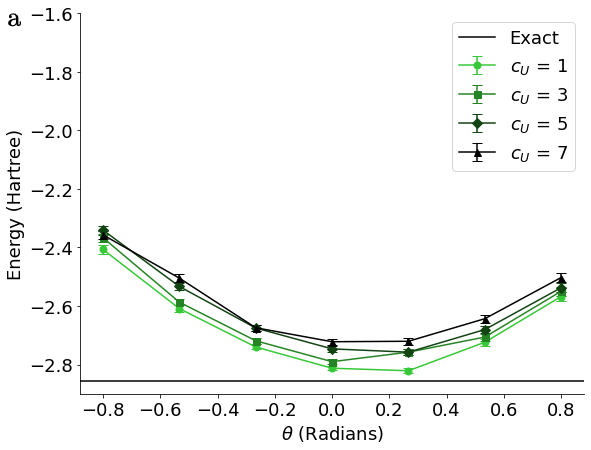}%
      \label{sfig:MS Sandwich Simulation}%
    }\hfill
    \subfloat{%
      \includegraphics[width=\columnwidth]{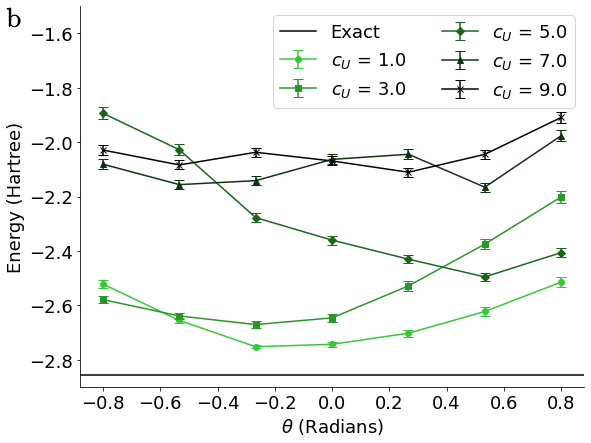}%
      \label{sfig:MS Sandwich Experiment}%
    }\hfill
    \caption{Comparison of simulation \textbf{(a)} and experimental \textbf{(b)} parameter sweep of the \textit{MS Before \& After} method. Gates are inserted before both MS gates in the unmitigated circuit. In simulation, the noise scales like the sum of the \textit{MS Before} and \textit{MS After} methods. In experiment, the noise scaling is less consistent, with a large outlier curve at $c_{U} = 5.0$.}
    \label{fig:MS Before and After}
\end{figure}

\begin{figure}
    \subfloat{%
      \includegraphics[width=\columnwidth]{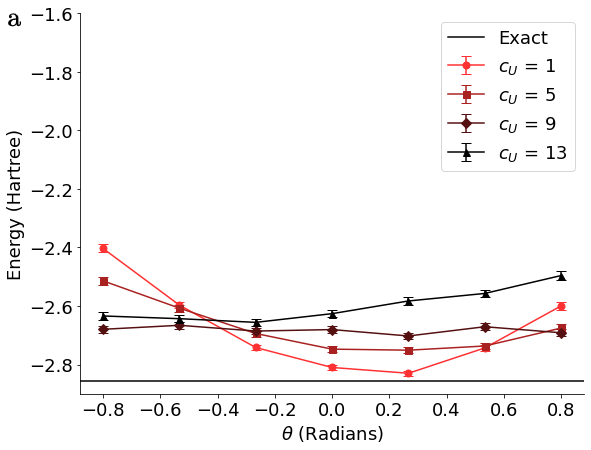}%
      \label{sfig:MS Four Simulation}%
    }\hfill
    \subfloat{%
      \includegraphics[width=\columnwidth]{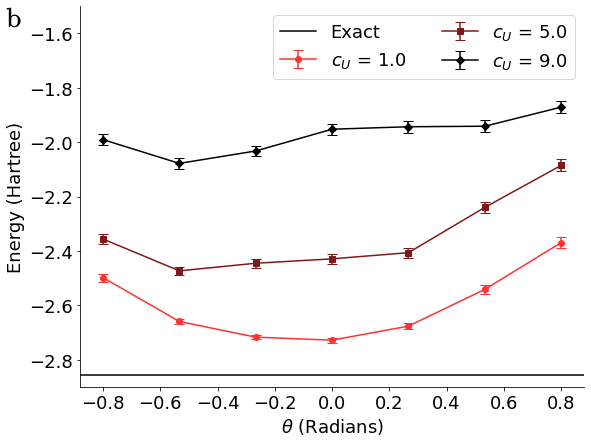}%
      \label{sfig:MS Four Experiment}%
    }\hfill
    \caption{Comparison of simulation \textbf{(a)} and experimental \textbf{(b)} parameter sweep of the \textit{MS Four} method. Four identical MS gates are inserted before each MS gate in the unmitigated circuit. The simulation results show that the energy landscape plateaus quickly as many gates are added to the circuit. In experiment, the results are more consistent, though the noise still accumulates rapidly. }
    \label{fig:MS Four}
\end{figure}

The \textit{MS Before} and \textit{MS After} methods (results shown in FIGs. \ref{sfig:MS Before Simulation} and \ref{sfig:MS After Simulation}) both show similar amounts of noise scaling, as they both insert the same number of two-qubit gates into the circuit. In simulation, the energy error incurred by this noise scaling is around $\sim 3 \%$ of the unmitigated circuit's minimum energy, which is in line with the results from \cite{Kandala_2019_noisy}, but still low. In the presence of the effect of finite sampling, some of these energy curves overlap, making extrapolation difficult. The only difference between the \textit{MS Before} and \textit{MS After} methods is where the gates are added, but the symmetry of the ansatz means that choice does not have an effect on the overall error. They maintain the parabolic shape of the energy landscape, and do not significantly shift the location of the minimum.

The experimental results for both the \textit{MS Before} and \textit{MS After} methods show substantially more energy error than simulation, upwards of $30\%$ of the unmitigated circuit's minimum energy at higher noise scale factors. This large energy gap makes extrapolation easy, though it limits the number of noisy energy estimates to $m=4$ before noise dominates. The energies still increase in accordance with the scale factor, though the parabolic shape of the energy landscape has plateaued somewhat. In the experimental data for the \textit{MS Before} method in FIG. \ref{sfig:MS Before Experiment}, the curve corresponding to $c = 3.0$ seems to be an outlier, exhibiting more noise than expected at negative $\theta$ and less noise at positive $\theta$. The data shown in FIG. \ref{sfig:MS After Experiment}, wherein we use the \textit{MS After} method, shows a similar, more consistent behavior.

For the \textit{MS Before \& After} method, we see the expected results in simulation. The noise scaling shown in FIG. \ref{sfig:MS Sandwich Simulation} is in line with what we would expect from the summation of FIGs. \ref{sfig:MS Before Simulation} and \ref{sfig:MS After Simulation}, sharing the same overall shape with more separation of the energy curves. Here, the maximum energy error of the most noise-scaled circuit compared to the unmitigated circuit is around $\sim 5 \%$, with the noisy energy curves overlapping less. Of all the simulated methods, this is the best for extrapolation.

The experimental results for the \textit{MS Before \& After} method shown in FIG. \ref{sfig:MS Sandwich Experiment} are less consistent as expected, particularly the $c = 5.0$ energy curve. Interestingly, we find that this experiment actually has less noise than either the \textit{MS Before} or \textit{MS After} methods. This may be because of better calibration between trials that reduces the effects of coherent errors.

Lastly, we have the \textit{MS Four} method, shown in FIG. \ref{fig:MS Four}. In simulation, the noise scaling is similar to that of the \textit{MS Before \& After} method, with overall noise scaling on the order of $\sim 7 \%$. However, the shape of the curves in FIG. \ref{sfig:MS Four Simulation} is different, with the $c = 9.0$ curve showing flattening we would normally expect from a state that has been greatly decohered. Interestingly, when we add yet more gates to the circuit, the $c = 13.0$ curve regains a slight parabolic structure. If we study the full energy landscape from $\theta = [-\pi, +\pi]$ and compare, we find that the behavior of the two noisiest energy curves is an artifact of zooming in on this particular region of $\theta$. Our noise model includes a $\sim 5\%$ over-rotation error in the angles of our $MS$ gates, which is dramatically amplified by this noise scaling method. This is good, insofar as we are capturing that kind of noise in our scaled circuit, but this coherent error in the noise model also has the effect of changing the location of the minimum energy.

This is in contrast with experiment, where we see less of a change in the shape of the energy landscape and more of a smooth increase in energy error, as is the case with the other gate insertion methods. The \textit{MS Four} method adds more noise to the circuit than any of the other gate insertion methods, on account of adding more gates, but still has an appropriate separation between the energy curves for extrapolation, as the data in FIG. \ref{sfig:MS Four Experiment} shows. The energy curves plateau at a similar rate to the \textit{MS Before \& After} experiment. However, using this technique, we are limited to only three insertions before the energy curve flattens, indicating too much noise in the circuit.

We initially assumed in our theoretical discussion in Section \ref{subsec:Gate Insertion Theory} that the gates would primarily be subject to a depolarizing channel. As a result, the noise scale factor for gate insertion methods would scale exponentially as a function of the total number of gates, particularly two-qubit gates, as we find they are the source of the majority of the noise in our experiments. For all the techniques outlined above, however, we find that this relationship does not hold true in simulation or experiment. Rather, for all the simulated and experimental results in this section, the relationship between the number of inserted gates and noise scaling is better modeled by a linear function. As such, we choose our extrapolation coefficients to simply be the ratio of the number of gates in the altered circuit over the number of gates in the unaltered circuit going forward.

\subsection{Gate Insertion Extrapolation}
\label{subsec:unitary_folding_exp}

All the gate insertion methods display enough of an energy gradient with increasing noise scaling for some measure of extrapolation. Of the four options, the two \textit{MS Before} and \textit{MS After} methods exhibit the best experimental results, insofar as they have the most data points, an obvious energy gradient, and the fewest outlier curves. We tested both methods with an in-depth extrapolation experiment, as will be described below, and found the results from the \textit{MS After} method to be most consistent. 

Using the same ansatz circuit and molecular parameters as in all other experiments, we performed a VQE optimization without any extrapolation. We used the COBYLA classical optimization algorithm \cite{SciPy} for our gradient descent, allocating 1000 samples per measurement for each iteration. The optimization converged to the minimum variational parameter after 21 steps.

\begin{figure}
    \includegraphics[width=\columnwidth]{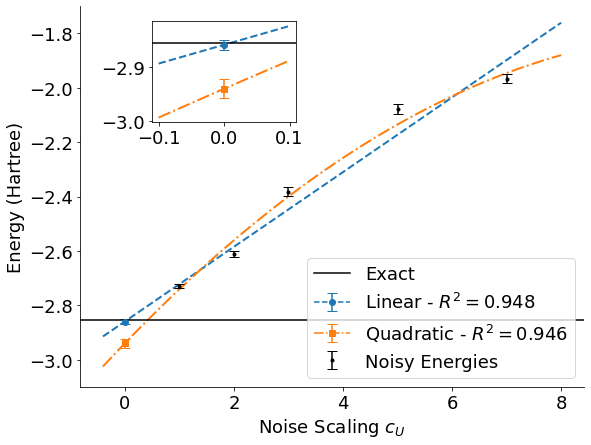}
    \caption{\label{fig:gate insertion extrapolation} \textbf{Hardware ---} Extrapolated energies using the \textit{MS After} method described in Section \ref{sec:gate insertion} and the \textit{first optimizing and then extrapolating} method described in Section \ref{sec:optimization}. The noisy energy estimates are plotted as a function of the noise scale factor $c_{U}$. The black data points represent the $m = 5$ noisy energy estimates made at $c_{U} = 1.0, 2.0, 3.0, 5.0, 7.0$ that were used for extrapolation. The data was fit using a least-squares method to a linear (blue dashed) and quadratic (orange dot-dashed) function. The linear fit results in a zero noise energy estimate of $-2.860 \pm 0.040$ Hartree, with $96.8 \%$ error suppression compared to the unaltered circuit. The quadratic fit results in an energy of $-2.940 \pm 0.017$ Hartree, with $34 \%$ error suppression compared to the unaltered circuit.}
\end{figure}

 We then performed extrapolation at that minimum parameter. The energy was estimated again using five noise scalings, again using 2000 samples per measurement. FIG. \ref{fig:gate insertion extrapolation} shows a clear increase in the energy due to the noise scaling. The noise scaled measurements were then extrapolated to two zero noise estimates, corresponding to a linear (blue dashed) and quadratic (orange dot-dashed) least-squares fit to the data. The linear fit falls outside the error bars of the data points, but describes the overall trend well, and results in a zero noise estimate very close to the true value. The quadratic fit has smaller residuals than the linear fit, as expected, but results in a zero noise estimate that overshoots the true value.
 
 The linear fit has an adjusted R-squared of 0.948 and a reduced chi-squared of 20, and the quadratic fit has an adjusted R-squared of 0.946 and a reduced chi-squared of 15. These values indicate that the two fits generally matches the data ($R^{2} \sim 1$), but that the error bars are too small to explain the residuals seen in the noisy energies ($\chi^{2} \gg 1)$. Our error bars capture the sampling error within each data point, but do not account for coherent errors such as qubit drift due to heating between experiments that would shift the noisy energy estimates.

These residuals (relative to the theoretical linear noise scaling) are enough to significantly skew the fits for higher order polynomials. We attempted Richardson extrapolation (fitting a polynomial of degree $m-1$ to $m$ data points) using this dataset, but found it overfit the data, resulting in zero noise energy estimates far from the ground state energy. The linear estimate falls within error bounds of the ground state estimate, though it has a slightly larger reduced chi-squared value than the quadratic estimate. These results are summarized in TABLE \ref{tab:extrapolation_results}.

\begin{table}
    \centering
    \begin{tabular}{|c|c|c|}
        \hline
        Fit & Linear & Quadratic \\
        \hline
        Absolute Error & $-0.004 \pm 0.009$ & $-0.085 \pm 0.017$ \\
        \hline
        Relative Error & $0.143 \% \pm 0.312 \%$ & $2.96 \% \pm 0.610 \% $ \\
        \hline
        Error & $96.8 \%$ & $34 \%$ \\
        Suppression & & \\
        \hline
    \end{tabular}
    \caption{Results of gate identity insertion based extrapolation. We report the absolute accuracy and precision in Hartree, the relative error normalized to the ground state energy, as well as the percentage error suppression due to ZNE.}
    \label{tab:extrapolation_results}
\end{table}


\section{Discussion}
\label{sec:discussion}

Our experimental and simulation results show that there is considerable nuance when it comes to implementing zero noise extrapolation techniques on a trapped-ion NISQ device. Such methods are promising because they do not require more qubits, however they are sensitive to the effects of finite sampling and necessitate a deeper understanding of the kinds of noise present in the system than we have access to with our noise model. A naive application of a given noise scaling method may not improve the accuracy or precision of an experiment by itself, due to the nature of the noise being a mismatch for the technique used.

We find that for our ion-trap system, QSCOUT, a time-stretching implementation of zero noise extrapolation does not scale the noise in the circuit enough to differentiate the resulting noisy energies in the presence of finite sampling. We contend this is true in general for qubits with long coherence times and little depolarizing noise, such that they maintain the prepared state over the time scales of lengthened gate operations. This also shows that different noise mitigation techniques can be a probe of the types of noise present in a device.

Additional experiments directly varying the detuning of the motional sidebands did a better job, but still failed to scale all the noise in the circuit. These experiments resulted in good fits to a linear or quadratic function, with bias similar to or better than no extrapolation. It is hard to say which result is more trustworthy, assuming no knowledge of the true ground state energy, as the theory supports the quadratic result and our experiment supports the linear result. If a large amount of error persists due to sideband detuning after device calibration, then this method may improve results, though for our device this was only a portion of our overall error.

We are also able to improve our results by scaling the noise in our circuits using gate-insertions, though the choice of specific gate insertion method impacts the effectiveness of error mitigation. For our particular device, with the majority of noise coming from the implementation of our two qubit gates and an absence of too many cancelling coherent errors, gate-based noise scaling works best. For devices with unknown error models or complicated relationships between gate parameters and noise, this method may be best, as it is more agnostic to noise source.

With a budget of 2000 samples for each Pauli measurement, we were able to show an improvement using a gate insertion method wherein the $MS$ gates in the second half of our ansatz circuit were scaled by repeated identity insertions of $(MS^{\dagger} MS)$. This method increased the noise in the ansatz circuit such that the resulting energy curves were distinguishable even in the presence of unmitigated coherent errors. We found that there were still additional sources of noise in our experimental data, most likely due to heating, that reduced the efficacy of extrapolation using higher order polynomials, though our linear fit obtained an improved zero noise energy estimate with $96.8 \%$ error suppression. Here we note that extrapolation is not variational, and therefore our estimate slightly overshoots the ground state.

There is some flexibility in how one chooses to integrate extrapolation into the classical optimization of VQE. We find that choosing to extrapolate the data first before optimizing, or optimizing the data first before extrapolation does not have a large effect on the resulting error in simulation. These different techniques all obtain energy estimates within $2 - 5 \%$ of the true energy minimum for a fixed sampling budget, with the difference between them diminishing with more samples. We found what matters most for the optimization is how many samples are used for each projective measurement at each iteration. For larger molecules with more complicated Hamiltonians, it may be necessary to instead perform extrapolation during the optimization routine, depending on the optimizer's susceptibility to noise.

We choose to only optimize the unstretched ansatz circuit, as we believe that the typical noise in the circuit does not meaningfully change the location of the optimal parameter for this problem. Limiting the optimization in this way allows us to allocate more samples to each iteration, improving the precision. Following this optimization, our extrapolated zero noise estimates for the expectation values of our UCCSD ansatz for \HeHplus\ lie closer to the true ground state energy. This improvement comes with a price: every zero noise energy estimate requires $m$ multiple noisy energy estimates, and thus $m$ times the number of samples for the same precision. If we were to allocate the same number of samples to the unmitigated circuit, we would see improvements to the precision of our estimates, at the price of reduced improvement to the accuracy.

We achieve a zero noise energy estimate with bias $-0.004$ and standard error $\pm 0.04$ Hartree, within error bounds of the ground state energy of the $\HeHplus\ $ molecule. These results are in line with previous VQE experiments. The original paper studying $\HeHplus\ $ \cite{peruzzo2014variational} estimated the ground state energy at $R = 0.8 \textup{\AA}$ with bias $\sim 0.05$ and standard error $\pm 0.05$ Hartree, relative to the exact value without manually correcting for errors, and bias $\sim -0.005$ and standard error $\pm 0.05$ Hartree after correction. Previous experiments \cite{Hempel_2018} using trapped ions studying $\Htwo\ $ instead of $\HeHplus\ $ similarly estimated the ground state energy at $R = 0.75 \textup{\AA}$ with bias $\sim 0.04$ and standard error $\pm 0.01$ Hartree, relative to the exact ground state energy. The original work \cite{Kandala_2019_noisy} applying Richardson Extrapolation to $\Htwo\ $ achieved a result with bias $\sim 0.005$ and standard error $\pm 0.01$ Hartree, relative to the ground state energy, bootstrapping a larger data set to reduce the effects of finite sampling.

Further research should be performed on larger systems to explore whether these findings hold true for more complex Hamiltonians and ansatz circuits. We believe that our results incorporating extrapolation into the optimization procedure inherent to VQE will change depending on the problem. For more complicated ansatz circuits with more optimization parameters, it may be necessary to perform some error mitigation during the optimization routine.

Zero noise extrapolation has been used to great effect recently on larger superconducting systems \cite{Tacchino_2020, Kim_2023a, Kim_2023b, Hour_2024}, using methods such as unitary folding. As system size increases, it may be necessary to use more sophisticated noise scaling methods such as probabilistic noise amplification to tackle multi-qubit errors like crosstalk. That said, as a trapped-ion device, the QSCOUT device experiences different types of noise than those superconducting devices, with our device only experiencing $\sim 0.17 \%$ crosstalk in practice \cite{clark2021engineering}. 

Our time-stretching results will be relevant for other trapped-ion systems, wherein depolarizing noise due to interactions with an environment is not the dominant source of noise. Our sideband detuning scaling results are similarly dependent on the degree of calibration of two-qubit MS gates. Lastly, our gate-insertion results depend less on the type of noise our gates experience, but may not scale certain coherent errors, and may be limited by circuit depth.

Careful consideration must be taken when choosing how many gates to insert into deeper circuits with many two-qubit gates, as it may not be feasible to scale every gate in the circuit without decohering the quantum state. This may require the use of selective gate insertion methods such as the \textit{MS Before} or \textit{MS After} methods outlined in this work. In general, these methods only multiplicatively scale the number of samples required to estimate the ground state energy, and therefore future experiments for larger molecules are more likely to be limited by the inherent resource scaling of VQE than by the overhead of error mitigation.

In order to realize larger applications of quantum computing on near-term devices, a better understanding of error mitigation techniques in the presence of finite sampling is needed. If current techniques fail on small scales, then they are unlikely to work on larger systems with deeper circuits and more complicated error models. How error mitigation techniques such as zero noise extrapolation interface with different qubit architectures, existing quantum algorithms, and other error mitigation techniques will be crucial going forward.

\section{Acknowledgements}
\label{sec:Acknowledgements}

The authors would like to thank Kenneth Rudinger and Andrew Baczewski from Sandia National Labs for their valuable comments and insight.

This project was funded by the U.S. Department of Energy, Office of Science, Office of Advanced Scientific Computing Research Quantum Testbed Program. Sandia National Laboratories is a multi-mission laboratory managed and operated by National Technology and Engineering Solutions of Sandia, LLC, a wholly owned subsidiary of Honeywell International Inc., for the U.S. Department of Energy’s National Nuclear Security Administration under contract DE-NA0003525.

%
\bibliographystyle{unsrtnat}
\bibliography{refs}


\appendix
\section{VQE Convergence}
\label{app:VQE Convergence}

\begin{figure}[b]
    \includegraphics[width=\columnwidth]{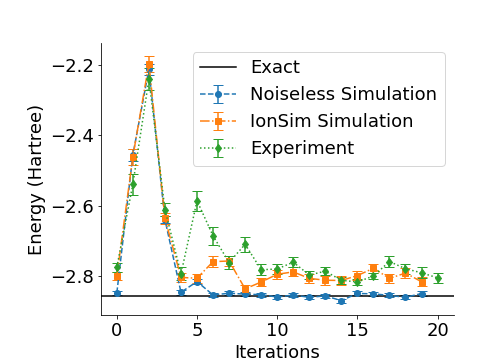}
    \caption{\textbf{Simulation and Experiment ---} Comparison of the COBYLA optimizer convergence in simulation and experiment. The noiseless simulation (blue, dashed), IonSim simulation (orange, dot-dashed), and experiment curves (green, dotted) all converge to an estimate in $\sim 10$ iterations. Without error mitigation, there is an energy error in the IonSim and experiment results.}
    \label{fig:Convergence Comparison}
\end{figure}

In Figure \ref{fig:Convergence Comparison}, we show plots of the convergence of our classical optimizer COBYLA running VQE for the HeH\textsuperscript{+} molecule at bond length $0.8$ \AA. All curves use 2000 samples to measure each term in the Hamiltonian, which is the same as we use in previous simulations and experiments. As discussed in Section \ref{sec:optimization}, we find it best to perform zero noise extrapolation after optimization, rather than at each iteration. These results show that both simulation and experiment are able to converge to a similar ground state estimate within a similar number of optimization iterations. There remains an energy error for these two noisy curves compared to the exact FCI energy shown in black.

\end{document}